\documentclass[aps,prresearch,preprint,groupedaddress,nofootinbib,
               longbibliography]{revtex4-2}

\usepackage{amsmath}
\usepackage{amssymb}
\usepackage{graphicx}
\usepackage[T1]{fontenc}
\usepackage[utf8]{inputenc}
\usepackage{booktabs}
\usepackage{array}[=2016-10-06]
\usepackage{tabularx}
\usepackage{float}   

\begin{document}

\title{Transitional evolution of the wave dispersion relation during helicon discharge ignition: an analytical theory from vacuum to steady operation}

\author{Lei Chang}
\email{leichang@cqu.edu.cn}
\email{leichang.plasma@gmail.com}
\affiliation{School of Electrical Engineering, Chongqing University, Chongqing 400044, China}

\date{\today}

\begin{abstract}
The dispersion relations of the helicon and Trivelpiece--Gould (TG) waves in a steady, fully developed magnetised plasma column are well established, but the discharge does not begin in that state. During ignition the electron density rises through five to seven orders of magnitude and the electron--neutral collision frequency falls by two to three, so the dispersion relation is itself a time-dependent object whose very existence must be justified. This paper develops a closed analytical theory of that transition for a cylindrical column in a uniform axial field, using a cold, collisional, two-species description that remains regular in the vacuum limit. Three structural results emerge. First, the vacuum limit is a degenerate double root of the quartic governing the perpendicular wavenumber; the helicon and TG branches are not independent modes that appear at some threshold but the two lobes of a single vacuum root, split by the gyrotropic and anisotropic parts of the dielectric tensor. The leading-order contribution to the discriminant cancels identically, so the branches separate linearly in density rather than with the square-root behaviour generic to a perturbed repeated root; the separation is correspondingly suppressed, and the helicon root remains pinned at its vacuum value until the electron density reaches about 10$^{15}$ m$^{-3}$ while the TG root departs immediately. Two thresholds acquire exact meaning: the TG root ceases to be radially evanescent precisely at \(n_{1} = \epsilon_{0}m_{e}\omega^{2}/e^{2}\), and the helicon root does so at \(n_{\text{cut}} = (4/\Lambda)\, n_{c}\). Second, the transient is organised by five characteristic densities and one magnetisation condition, spaced by powers of \(\Lambda = \omega_{\text{ce}}/\omega\) and crossed in a fixed order; a compact criterion, \(n_{\text{res}}/n_{c} = (4/\Lambda)\sqrt{1 + T^{2}/k_{z}^{2}}\), determines whether a bounded eigenmode resonates before or after the helicon/TG branches coalesce, and shows that no bounded mode resonates before coalescence unless \(\omega_{\text{ce}} > 4\omega\). Third, a two-timescale analysis separates local from global adiabaticity: the instantaneous dielectric tensor is accurate to better than \(10^{- 3}\) throughout, whereas the driven boundary-value problem is strongly non-adiabatic during the first tens of microseconds, where the cavity fill time is comparable to the density e-folding time. All non-adiabatic physics of ignition therefore resides in the global problem rather than in the constitutive law. Explicit predictions and benchmark values are given for both numerical and experimental verifications.
\end{abstract}

\maketitle

\section{Introduction}
Helicon sources produce high plasma densities at modest magnetic field and modest radiofrequency power. The discharge originates in the standing-wave plasma production reported by Boswell \cite{r1} and in the efficient generation by whistler waves near the lower-hybrid frequency that followed \cite{r2}, confirmed and extended by Komori and co-workers \cite{r3}; the first three decades of that development are surveyed by Boswell and Chen \cite{r4,r5}, and the state of the field since then in the reviews \cite{r6,r7,r8,r9,r10}. The reason for the efficiency has been understood since the work of Chen \cite{r11} and of Shamrai and Taranov \cite{r12} to lie in the coupling between two branches of the bounded cold-plasma dispersion relation. The helicon branch is a weakly damped whistler-like wave that fills the column volume; the Trivelpiece--Gould (TG) branch is a short-wavelength, quasi-electrostatic wave confined to a thin edge layer and damped rapidly by collisions \cite{r13,r14}. Antenna power couples preferentially to the helicon branch, converts to the TG branch near the boundary, and is absorbed there \cite{r15,r16,r17}. The theory of this coupling in a steady, uniform or weakly non-uniform column is mature, and its predictions for wavenumbers, radial eigenmodes and absorption profiles are routinely compared with experiment. On the strength of that efficiency the discharge now serves applications as varied as materials processing \cite{r18}, electrodeless spacecraft propulsion and the magnetic nozzles that accompany it \cite{r19,r20,r21}, and non-inductive current drive in tokamaks \cite{r22,r23}, so that its start-up behaviour is of interest well beyond any one device.

The discharge does not, however, begin in that state. A helicon source starts as an evacuated dielectric tube containing neutral gas and a residual electron population of essentially vacuum density. Over an interval of tens to hundreds of microseconds it passes through a capacitive (E) phase, an inductive (H) phase, and finally jumps into the wave-coupled (W) phase \cite{r24,r25}. Across that interval the electron density rises through five to seven orders of magnitude; the electron--neutral collision frequency falls by two to three orders of magnitude as the gas is heated and depleted; and the radial density profile reorganises from a nearly flat, edge-dominated shape into a centrally peaked, diffusion-controlled one. Every one of these changes acts directly on the dielectric tensor, and hence on the dispersion relation.

Existing treatments of the ignition transient are overwhelmingly numerical and transport-oriented: global or two-dimensional models resolve the particle and energy balance and thereby capture the E--H--W jumps and their hysteresis, with the wave field obtained from a frequency-domain solve at each time step \cite{r18,r26,r27,r28,r29,r30}. Such models reproduce the phenomenology but do not answer the structural questions. In what sense does a dispersion relation exist at all while the medium is changing? How does the relation deform continuously from the vacuum relation \(\omega = ck\) into the coupled helicon--TG pair, rather than appearing discontinuously at some threshold? Where, i.e. in density, in radius and in time, does the notion of an instantaneous dispersion relation cease to be meaningful? These questions are analytical, and answering them is a precondition for interpreting the numerical results, not a supplement to them, and of great value for experimental design and optimisations.

The present paper develops such a theory. The plasma is described by a cold, collisional, two-species fluid model whose dielectric tensor reduces smoothly to the identity as the density vanishes, so that the vacuum state is contained within the same formalism as the developed discharge. Section 3 shows that the vacuum limit of the quartic governing the perpendicular wavenumber is a degenerate double root, and that plasma formation acts as a degeneracy-breaking perturbation controlled by the gyrotropic element and by the anisotropy; the helicon and TG branches are therefore the two lobes of a single vacuum root. Section 4 treats the bounded problem, i.e. combining the resonance condition of the cylindrical eigenmode with the coalescence condition of Shamrai and Taranov, and yields the compact ratio \(n_{\text{res}}/n_{c} = (4/\Lambda)\sqrt{1 + T^{2}/k_{z}^{2}}\), in which \(T\) is the perpendicular eigenvalue of the radial boundary-value problem and \(\Lambda = \omega_{\text{ce}}/\omega\) the magnetisation ratio; it separates bounded modes that resonate cleanly from those whose resonance is embedded in the coalesced regime, and implies a lower bound \(\omega_{\text{ce}} > 4\omega\) on the field required for clean helicon operation, i.e. independent of density, power and gas. Section 5 formulates the transient as a two-timescale problem and distinguishes local from global adiabaticity; the correction to the instantaneous constitutive relation is shown to be of relative order \((\omega\tau_{n})^{- 1}\) and is negligible throughout; the driven boundary-value problem, by contrast, has a response time \(2Q/\omega\) that becomes comparable to the density e-folding time near a resonance. Sections 6 and 7 close the loop with the transport problem and with power deposition, and Section 8 states the consequences for numerical modelling and for experiment in a form that can be tested directly. Throughout, every closed-form result is verified against numerical evaluation of the full quartic; the verification residuals are reported in Section 8.4.

\section{Physical model and ordering}
\subsection{Configuration}
A complete list of symbols is collected in Appendix D. A cylindrical plasma column of radius \(a\) occupies a uniform axial magnetic field \(B_{0}\widehat{z}\) and is driven by an external antenna at fixed angular frequency \(\omega\). Fields are decomposed into azimuthal mode number \(m\) and axial wavenumber \(k_{z}\),
\begin{equation}
\mathbf{E}(r,\theta,z,t) = \widetilde{\mathbf{E}}(r,t)\,\text{exp}\left\lbrack i\left( m\theta + k_{z}z - \omega t \right) \right\rbrack
\end{equation}
with the slow evolution of the discharge carried entirely by the envelope \(\widetilde{\mathbf{E}}(r,t)\). The electron density \(n_{e}(r,t)\) and momentum-transfer collision frequency \(\nu_{m}(r,t)\) are supplied by a transport model (Section 6) and are treated as slowly varying parameters in Sections 3 and 4.

All numerical illustrations use the reference case of Table 1, chosen to represent a laboratory helicon source operating at low magnetic field, where the helicon and TG branches interact most strongly. Note that \(k_{z} \gg k_{0} = \omega/c\): the antenna drives a slow wave, which in vacuum is a purely evanescent near field. The subject of this paper is how that evanescent near field becomes a propagating helicon.
\begin{table}[htbp]
\centering
\footnotesize
\setlength{\tabcolsep}{3pt}
\caption{Reference case used for all numerical illustrations.}
\label{tab:1}
\begin{tabularx}{\textwidth}{>{\raggedright\arraybackslash}X>{\raggedright\arraybackslash}X>{\raggedright\arraybackslash}X}
\toprule
Quantity & Symbol & Value \\
\midrule
Drive frequency & \(f = \omega/2\pi\) & 13.56 MHz \\
Axial magnetic field & \(B_{0}\) & 100 G \\
Tube radius & \(a\) & 2.5 cm \\
Source length & \(L\) & 0.4 m \\
Working gas & --- & argon \\
Azimuthal mode number & \(m\) & 1 \\
Axial wavenumber & \(k_{z}\) & 31.4 m$^{-1}$ (\(\lambda_{z}\) = 20 cm) \\
Electron gyrofrequency & \(\omega_{\text{ce}}\) & 1.759 $\times$ 10$^{9}$ s$^{-1}$ \\
Ion gyrofrequency & \(\omega_{\text{ci}}\) & 2.412 $\times$ 10$^{4}$ s$^{-1}$ \\
Magnetisation ratio & \(\Lambda = \omega_{\text{ce}}/\omega\) & 20.64 \\
Vacuum wavenumber & \(k_{0} = \omega/c\) & 0.284 m$^{-1}$ \\
\bottomrule
\end{tabularx}
\end{table}

\subsection{Ordering parameters}
Six dimensionless groups control the problem, and their relative magnitude defines the stage of the discharge:
\begin{equation}
\Lambda = \frac{\omega_{\text{ce}}}{\omega},\quad\quad\Lambda_{\nu} = \frac{\omega_{\text{ce}}}{\nu_{m}},\quad\quad\delta_{\nu} = \frac{\nu_{m}}{\omega}
\end{equation}
\begin{equation}
\epsilon_{\text{loc}} = \frac{1}{\omega}\left| \frac{{\dot{n}}_{e}}{n_{e}} \right|,\quad\quad\epsilon_{\text{glob}} = \tau_{\text{cav}}\left| \frac{{\dot{n}}_{e}}{n_{e}} \right|,\quad\quad\epsilon_{r} = \frac{1}{k_{\bot}L_{n}}
\end{equation}
Here \(\omega_{\text{ce}} = eB_{0}/m_{e}\) is the electron gyrofrequency, and an overdot denotes \(\partial/\partial t\). The first three groups characterise the medium. \(\Lambda\) is the magnetisation ratio, the number of electron gyro-orbits per wave period; it is large whenever the wave is a whistler rather than a light wave, and it sets the spacing of the characteristic densities of Section 3.4. \(\Lambda_{\nu}\) is the magnetisation parameter, the number of gyro-orbits an electron completes between momentum-transfer collisions; \(\Lambda_{\nu} > 1\) is the condition for the electrons to be magnetised at all, and Section 3.5 shows that it, rather than any density, is what permits the helicon and TG branches to exist as separate waves. \(\delta_{\nu}\) is the collisionality, i.e. the collision frequency measured in units of the drive frequency; it is the small parameter of the collisionless limit and controls the damping of both branches. During ignition \(\Lambda\) is fixed by the applied field, whereas \(\Lambda_{\nu}\) rises and \(\delta_{\nu}\) falls by two to three orders of magnitude as the gas is heated and depleted, so the medium is demagnetised and strongly collisional at breakdown and magnetised and weakly collisional at steady state.

The remaining three groups characterise the transient itself. \(L_{n} = \left| \nabla\text{ln}n_{e} \right|^{- 1}\) is the density scale length and \(\tau_{\text{cav}} = 2Q/\omega\) the response time of the driven cavity mode, \(Q\) being its quality factor, so that \(\epsilon_{\text{loc}}\) compares the density e-folding time with the wave period and \(\epsilon_{\text{glob}}\) compares it with the time the cavity needs to fill. The pair \((\epsilon_{\text{loc}},\epsilon_{\text{glob}})\) carries the entire non-adiabatic content of the problem and is analysed in Section 5; \(\epsilon_{r}\) governs the validity of the radial Wentzel--Kramers--Brillouin (WKB) description and measures the perpendicular wavelength against the density scale length, and is small for the TG branch but only marginally so for the helicon branch.

\subsection{Constitutive relation for an evolving electron population}
Because the electron population is being created during the transient, the constitutive relation requires care. Starting from the momentum density rather than the velocity,
\begin{equation}
\frac{\partial}{\partial t}\left( n_{e}m_{e}\mathbf{v}_{e} \right) = - en_{e}\left( \mathbf{E} + \mathbf{v}_{e} \times \mathbf{B}_{0} \right) - n_{e}m_{e}\nu_{m}\mathbf{v}_{e} + S_{\text{iz}}m_{e}\mathbf{v}_{\text{birth}} - S_{\text{loss}}m_{e}\mathbf{v}_{e}
\end{equation}
where \(S_{\text{iz}}\) is the ionisation source, \(S_{\text{loss}}\) the particle loss rate and \(\mathbf{v}_{\text{birth}}\) the mean velocity of newly created electrons. Electrons produced by ionisation of a cold neutral background are born with no radiofrequency drift, so \(\mathbf{v}_{\text{birth}} = 0\) and that term vanishes. Writing \(\mathbf{J} = - en_{e}\mathbf{v}_{e}\) and \(\mathbf{\Omega} = e\mathbf{B}_{0}/m_{e}\) gives
\begin{equation}
\frac{\partial\mathbf{J}}{\partial t} + \nu_{\text{eff}}\mathbf{J} + \mathbf{J} \times \mathbf{\Omega} = \epsilon_{0}\,\omega_{\text{pe}}^{2}(t)\,\mathbf{E},\quad\quad\nu_{\text{eff}} = \nu_{m} + \frac{S_{\text{loss}}}{n_{e}}
\end{equation}

Two consequences follow. Ionisation does not damp the radiofrequency current: newly born electrons dilute the mean velocity but carry no current, so \(\mathbf{J}\) is continuous across a birth event and no term in \({\dot{n}}_{e}/n_{e}\) appears. This is worth stating explicitly because such a term is easily introduced in error when the momentum equation is written for \(\mathbf{v}_{e}\) rather than for \(n_{e}m_{e}\mathbf{v}_{e}\). Particle loss, by contrast, does damp the current, since electrons lost to the wall carry their radiofrequency momentum with them; in quasi-steady state \(S_{\text{loss}} \simeq S_{\text{iz}}\) and \(\nu_{\text{eff}} = \nu_{m} + \nu_{\text{iz}}\). For the reference case \(\nu_{\text{iz}} \sim 10^{4}\)--\(10^{5}\) s$^{-1}$ is negligible beside \(\nu_{m} \sim 10^{7}\)--\(10^{10}\) s$^{-1}$, but the term is the only route by which the transport model enters the constitutive law directly, and it is retained here for completeness.

\subsection{Dielectric tensor and its vacuum limit}
Taking \(\partial_{t} \rightarrow - i\omega\) (the correction to this step is quantified in Section 5.3) and adding the ion species yields the cold collisional tensor in Stix form \cite{r31},
\begin{equation}
\mathbf{\varepsilon} = \begin{pmatrix}
S & - iD & 0 \\
\text{iD} & S & 0 \\
0 & 0 & P \\
\end{pmatrix}
\end{equation}
where the sum runs over species \(s\), with squared plasma frequencies \(\omega_{\text{p}s}^{2} = n_{s}q_{s}^{2}/(\epsilon_{0}m_{s})\), signed gyrofrequencies \(\Omega_{s} = q_{s}B_{0}/m_{s}\), and collisionally shifted frequencies \({\widetilde{\omega}}_{s} = \omega + i\nu_{s}\), \(\nu_{s}\) being the momentum-transfer collision frequency of species \(s\):
\begin{equation}
S = 1 - \sum_{s}^{}\frac{\omega_{\text{p}s}^{2}\,{\widetilde{\omega}}_{s}}{\omega\left( {\widetilde{\omega}}_{s}^{2} - \Omega_{s}^{2} \right)},\quad\quad D = \sum_{s}^{}\frac{\omega_{\text{p}s}^{2}\,\Omega_{s}}{\omega\left( {\widetilde{\omega}}_{s}^{2} - \Omega_{s}^{2} \right)},\quad\quad P = 1 - \sum_{s}^{}\frac{\omega_{\text{p}s}^{2}}{\omega \,{\widetilde{\omega}}_{s}}
\end{equation}
The property essential to the present work is the regularity of the vacuum limit: as \(n_{e} \rightarrow 0\), every \(\omega_{\text{p}s}^{2} \rightarrow 0\) and
\begin{equation}
S \rightarrow 1,\quad\quad D \rightarrow 0,\quad\quad P \rightarrow 1,\quad\quad\mathbf{\varepsilon} \rightarrow \mathbb{I}
\end{equation}
The tensor deforms smoothly from the identity, which is precisely what permits the dispersion relation to be followed continuously from vacuum. Reduced descriptions that discard the displacement current, e.g. including the EMHD model used almost universally in helicon theory, lose this property and become singular at ignition, as shown in Section 3.6.

\section{Local dispersion relation and its deformation from vacuum}
\subsection{The quartic and its invariants}
Substituting a plane wave into \(\nabla \times \nabla \times \mathbf{E} = k_{0}^{2}\,\mathbf{\varepsilon} \cdot \mathbf{E}\) and eliminating the polarisation gives a quadratic in \(k_{\bot}^{2}\) (Appendix A),
\begin{equation}
\mathcal{A}\, k_{\bot}^{4} + \mathcal{B}\, k_{\bot}^{2} + \mathcal{C} = 0
\end{equation}
\begin{equation}
\mathcal{A} = S,\quad\quad\mathcal{B} = (S + P)\left( k_{z}^{2} - Sk_{0}^{2} \right) + D^{2}k_{0}^{2},\quad\quad\mathcal{C} = P\left\lbrack \left( k_{z}^{2} - Sk_{0}^{2} \right)^{2} - D^{2}k_{0}^{4} \right\rbrack
\end{equation}
The two roots follow from the quadratic formula, but for tracking the transient the elementary invariants are more useful, being polynomial in the plasma parameters and hence free of branch-cut ambiguity:
\begin{equation}
k_{\bot +}^{2} + k_{\bot -}^{2} = - \frac{\mathcal{B}}{\mathcal{A}},\quad\quad k_{\bot +}^{2}\, k_{\bot -}^{2} = \frac{\mathcal{C}}{\mathcal{A}} = \frac{P\left\lbrack \left( k_{z}^{2} - Sk_{0}^{2} \right)^{2} - D^{2}k_{0}^{4} \right\rbrack}{S}
\end{equation}

\subsection{The vacuum limit is a degenerate double root}
Setting \(S = P = 1\) and \(D = 0\) reduces the quartic to
\begin{equation}
k_{\bot}^{4} + 2\left( k_{z}^{2} - k_{0}^{2} \right)k_{\bot}^{2} + \left( k_{z}^{2} - k_{0}^{2} \right)^{2} = \left( k_{\bot}^{2} + k_{z}^{2} - k_{0}^{2} \right)^{2} = 0
\end{equation}
The discriminant vanishes identically: the vacuum limit is a repeated root \(k_{\bot}^{2} = k_{0}^{2} - k_{z}^{2}\), which for the antenna-driven slow wave \(k_{z} > k_{0}\) is negative,
\begin{equation}
k_{\bot} = i\kappa,\quad\quad\kappa = \sqrt{k_{z}^{2} - k_{0}^{2}} \simeq k_{z}
\end{equation}
The vacuum field is the evanescent near field \(I_{m}(\kappa r)\), with no propagation and no energy transport into the volume. For the reference case \(\kappa = 31.4\) m$^{-1}$ and \(\kappa a = 0.79\), so the vacuum profile is close to the quasi-static \(r^{m}\) form; for \(m = 1\) this is the linear \(\left| B_{z} \right| \propto r\) visible in the first panel of Figure 7.

The physical content of the degeneracy is that the two circular polarisations are indistinguishable in vacuum: there is no birefringence. Plasma formation breaks the degeneracy, and it does so through the gyrotropic element \(D\) and the anisotropy \(P - S\). The helicon and TG branches are thus not independent objects that appear at some threshold density; they are the two lobes of a single vacuum root.

\subsection{Degeneracy breaking near ignition}
Writing \(S = 1 + \delta S\), \(P = 1 + \delta P\), \(D = \delta D\) with \(\delta S\), \(\delta P\), \(\delta D\) small, and \(\mathcal{K} \equiv k_{z}^{2} - k_{0}^{2} > 0\), expansion of the discriminant shows that its leading term cancels identically, leaving
\begin{equation}
\mathcal{B}^{2} - 4\mathcal{A}\mathcal{C} = \left( \delta P - \delta S \right)^{2}\mathcal{K}^{2} + 4\,\delta D^{2}k_{0}^{2}k_{z}^{2} + O(3) \simeq \left( P - S \right)^{2}k_{z}^{4} + 4D^{2}k_{0}^{2}k_{z}^{2}
\end{equation}
so that
\begin{equation}
k_{\bot \pm}^{2} \simeq - \mathcal{K}\left( 1 + \frac{\delta P - \delta S}{2} \right) + \delta S\,k_{0}^{2} \pm \frac{1}{2}\sqrt{\left( \delta P - \delta S \right)^{2}k_{z}^{4} + 4\,\delta D^{2}k_{0}^{2}k_{z}^{2}}
\end{equation}

Three consequences follow. First, the splitting is controlled by the anisotropy \(P - S\) and the gyrotropy \(D\), not by the mean permittivity: a dense but unmagnetised plasma, for which \(\delta D = 0\) and \(\delta P = \delta S\), never splits the root at all and merely deepens the evanescence. Magnetisation is what creates two branches. Second, the degeneracy is broken more weakly than the general theory of perturbed repeated roots would suggest. A double root of a quadratic subjected to a perturbation of size \(\zeta\) normally splits as \(\zeta^{1/2}\). Here the term of first order in \(n_{e}\) cancels identically between \(\mathcal{B}^{2}\) and \(4\mathcal{AC}\), leaving a discriminant quadratic in \(n_{e}\), since \(P - S\) and \(D\) are each linear in \(n_{e}\). Its square root is therefore linear, and so is the separation:
\begin{equation}
k_{\bot +}^{2} - k_{\bot -}^{2} \propto n_{e},\quad\quad\delta k_{\bot} = \frac{k_{\bot +}^{2} - k_{\bot -}^{2}}{k_{\bot +} + k_{\bot -}} \propto n_{e}
\end{equation}
the denominator remaining finite at \(2i\kappa \) in the vacuum limit. Direct evaluation of the quartic confirms the exponent: the separation grows by exactly a factor of ten for every tenfold rise in density between \(n_{e} = 10^{8}\) and \(10^{12}\) m$^{-3}$. The practical consequence is that the branches emerge slowly, and that the earliest phase of the discharge is far less sensitive to small density changes than a naive degeneracy argument would predict. Third, equating the two contributions to the discriminant locates the crossover from anisotropy-dominated to gyrotropy-dominated splitting. With \(\left| \delta P \right| \gg \left| \delta S \right|\), this gives \(k_{z}^{2}c^{2} = 4\omega^{4}/\omega_{\text{ce}}^{2}\), a condition on the antenna rather than on the plasma. For the reference case, we have \(k_{z}^{2}c^{2} = 8.9 \times 10^{19}\) s$^{-2}$ against \(4\omega^{4}/\omega_{\text{ce}}^{2} = 6.8 \times 10^{13}\) s$^{-2}$, so anisotropic splitting dominates by six orders of magnitude and the earliest phase of the discharge is insensitive to polarisation.

\subsection{Characteristic densities and the \(\Lambda\) ladder}
Each element of the tensor departs from its vacuum value at a different density. Requiring each departure to be of order unity defines a natural ladder,
\begin{equation}
\begin{gathered}
n_{1}:\mspace{6mu}\omega_{\text{pe}} = \omega\mspace{6mu} \Rightarrow \mspace{6mu} n_{1} = \frac{\epsilon_{0}m_{e}\omega^{2}}{e^{2}}, \\
n_{2}:\mspace{6mu}\omega_{\text{pe}}^{2} = \omega\omega_{\text{ce}}\mspace{6mu} \Rightarrow \mspace{6mu} n_{2} = n_{1}\Lambda, \\
n_{3}:\mspace{6mu}\omega_{\text{pe}} = \omega_{\text{ce}}\mspace{6mu} \Rightarrow \mspace{6mu} n_{3} = n_{1}\Lambda^{2}
\end{gathered}
\end{equation}
so that
\begin{equation}
n_{1}:n_{2}:n_{3} = 1:\Lambda:\Lambda^{2}
\end{equation}

Two of these thresholds are not merely order-of-magnitude markers but have exact meaning, which emerges on asking where each root ceases to be radially evanescent, that is where \(\left| \beta \right|\) crosses \(k_{z}\), \(\beta = (k_{\bot}^{2} + k_{z}^{2})^{1/2}\) denoting the total wavenumber of a branch. For the upper root \(\beta_{+}^{2} = k_{z}^{2}\left( 1 - P/S \right)\), so with \(S \simeq 1\) at low density the crossing occurs precisely at \(P = 0\):
\begin{equation}
\left| \beta_{+} \right| = k_{z}\quad \Leftrightarrow \quad n_{e} = n_{1}
\end{equation}
Numerical solution of the full quartic places the crossing at \(2.2808 \times 10^{12}\) m$^{-3}$ against \(n_{1} = 2.2808 \times 10^{12}\) m$^{-3}$, agreeing to five figures. The lower root, by contrast, remains pinned at the vacuum value \(\beta = k_{0}\) to within 1\% until \(n_{e} = 8.3 \times 10^{14}\) m$^{-3}$, that is until \(n_{e} \simeq n_{3}\); and it crosses \(k_{z}\) only at the helicon cutoff density obtained from \(\beta k_{z} = \omega\mu_{0}n_{e}e/B_{0}\) with \(\beta = k_{z}\),
\begin{equation}
n_{\text{cut}} = \frac{k_{z}^{2}B_{0}}{\omega\mu_{0}e} = \frac{4}{\Lambda}\, n_{c}
\end{equation}
which evaluates to \(5.75 \times 10^{17}\) m$^{-3}$ for the reference case; the full quartic places the crossing at \(5.48 \times 10^{17}\) m$^{-3}$, 5\% lower, the difference being the displacement-current correction that the EMHD relation omits. The thresholds therefore mark distinct physical events rather than a single graded transition: \(n_{1}\) is where the TG branch begins to propagate radially, \(n_{3}\) is where the helicon branch first leaves its vacuum value, and \(n_{\text{cut}}\) is where the helicon branch itself begins to propagate. The TG branch is thus established some five orders of magnitude in density before the helicon branch exists at all.

The ignition transient climbs this ladder in order, with rungs spaced by a factor \(\Lambda\), i.e. for the reference case a factor of 20.6, or about 1.3 orders of magnitude. Table 2 lists the values together with the coalescence density \(n_{c}\) derived in Section 3.7. The first three are properties of the medium alone; the fourth involves \(k_{z}\) and is therefore a joint property of medium and antenna. The gap of nearly 3.5 orders of magnitude between \(n_{3}\) and \(n_{c}\) spans the entire operating region of a conventional inductive discharge, and it is there that the TG branch exists as a well-separated, strongly damped edge wave.
\begin{table}[htbp]
\centering
\footnotesize
\setlength{\tabcolsep}{3pt}
\caption{Characteristic densities of the ignition transient, with their defining conditions, physical meaning and values for the reference case.}
\label{tab:2}
\begin{tabularx}{\textwidth}{ll>{\raggedright\arraybackslash\hsize=1.000\hsize}Xl}
\toprule
Threshold & Condition & Physical meaning & Value \\
\midrule
\(n_{1}\) & \(\omega_{\text{pe}} = \omega\), \(P = 0\) & TG root crosses \(k_{z}\); TG branch becomes radially propagating & 2.28 $\times$ 10$^{12}$ m$^{-3}$ \\
\(n_{2}\) & \(\omega_{\text{pe}}^{2} = \omega\omega_{\text{ce}}\), \(\left| D \right| = 1\) & Gyrotropy becomes order unity; birefringence appears & 4.71 $\times$ 10$^{13}$ m$^{-3}$ \\
\(n_{3}\) & \(\omega_{\text{pe}} = \omega_{\text{ce}}\), \(S = 2\) & Helicon root leaves its vacuum value \(\beta = k_{0}\) & 9.72 $\times$ 10$^{14}$ m$^{-3}$ \\
\(n_{\text{cut}}\) & \(\beta_{-} = k_{z}\) & Helicon branch becomes radially propagating & 5.75 $\times$ 10$^{17}$ m$^{-3}$ \\
\(n_{c}\) & \(4k_{s}^{2} = k_{w}^{2}\) & Helicon and TG branches coalesce & 2.97 $\times$ 10$^{18}$ m$^{-3}$ \\
\bottomrule
\end{tabularx}
\end{table}

\subsection{Collisional demagnetisation}
At ignition the neutral pressure is high and the electron--neutral collision frequency correspondingly large. For argon at 10--100 mTorr with \(T_{e} \sim 2\)--4 eV, \(\nu_{m} \simeq n_{g}\left\langle \sigma v \right\rangle\) with \(\left\langle \sigma v \right\rangle \sim 10^{- 13}\) m$^{3}$ s$^{-1}$ gives \(\nu_{m} \sim 3 \times 10^{7}\)--\(3 \times 10^{8}\) s$^{-1}$, and values approaching \(10^{10}\) s$^{-1}$ occur in the high-pressure fill used to ensure reliable breakdown. These are to be compared with \(\omega_{\text{ce}} = 1.76 \times 10^{9}\) s$^{-1}$. When \(\nu_{m} \gtrsim \omega_{\text{ce}}\) the electron cannot complete a gyro-orbit between collisions (here and in what follows $\nu$ denotes the effective momentum-transfer rate $\nu_{\text{eff}}$ of Eq. (5)) and the gyrotropic response is destroyed,
\begin{equation}
D = \frac{\omega_{\text{pe}}^{2}\,\Omega_{e}}{\omega\left\lbrack \left( \omega + i\nu \right)^{2} - \Omega_{e}^{2} \right\rbrack}\mspace{6mu} \rightarrow \mspace{6mu} + \frac{\omega_{\text{pe}}^{2}\,\omega_{\text{ce}}}{\omega \,\nu^{2}}\mspace{6mu} \rightarrow \mspace{6mu} 0\text{\quad\quad}\left( \nu \gg \omega_{\text{ce}} \right)
\end{equation}
so the plasma is effectively unmagnetised however large \(B_{0}\) may be. Since \(D\) is what breaks the vacuum degeneracy, the conclusion is sharp: separated helicon and TG branches cannot exist until \(\nu_{m}\) has fallen below \(\omega_{\text{ce}}\), irrespective of density. The magnetisation threshold \(\Lambda_{\nu} = 1\) is crossed in time by gas heating and depletion, not by ionisation, and this is why the E--H--W sequence is a temporal sequence and not merely a density sequence. Figure 1 shows the consequence directly: below \(n_{e} \sim 10^{15}\) m$^{-3}$ neither root is a useful propagating wave, though for different reasons. The lower root has \(\left| \beta \right| \simeq k_{0} \ll k_{z}\) and is radially evanescent while remaining almost undamped; the upper root propagates radially above \(n_{1}\) but, at the collision frequency prevailing there, carries \(\left| \text{Im}\,\beta  \right|\) comparable to or larger than \(\text{Re}\,\beta \) and is destroyed as fast as it propagates. At \(n_{e} = 10^{14}\) m$^{-3}$ with \(\nu_{m} = 3 \times 10^{9}\) s$^{-1}$ the ratio is 2.1; the fixed \(\nu = 10^{7}\) s$^{-1}$ of Figure 1, chosen there to isolate the density dependence, gives 0.06.
\begin{figure}[htbp]
\centering
\includegraphics[width=0.95\linewidth]{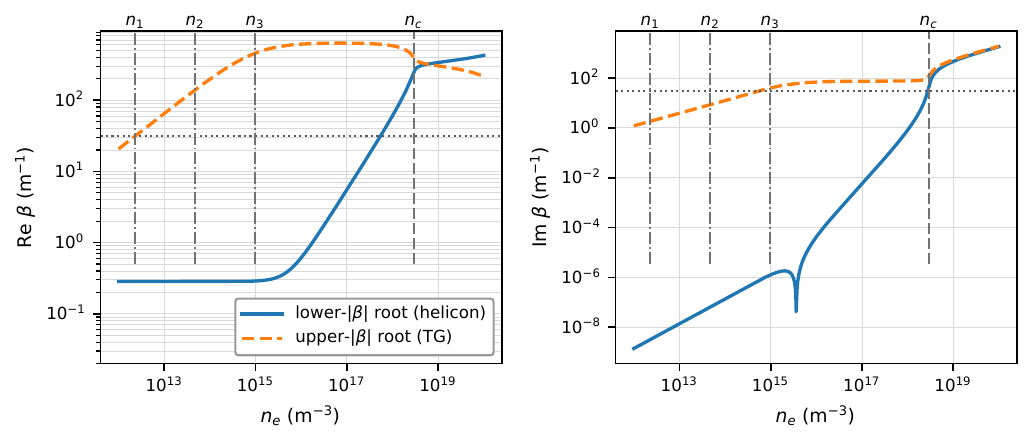}
\caption{Evolution of the two branches of the local dispersion relation as the electron density rises from near-vacuum to beyond helicon operating density. Left: real part of the total wavenumber \(\beta = (k_{\bot}^{2} + k_{z}^{2})^{1/2}\). Right: imaginary part. Dashed vertical lines mark the threshold densities \(n_{1}\), \(n_{2}\), \(n_{3}\) and the coalescence density \(n_{c}\); the dotted horizontal line is \(k_{z}\).}
\label{fig:1}
\end{figure}

\subsection{Electron-magnetohydrodynamic reduction and its failure at ignition}
For \(\omega_{\text{ci}} \ll \omega \ll \omega_{\text{ce}} \ll \omega_{\text{pe}}\) with immobile ions and negligible displacement current, the problem collapses to a far more transparent form. The generalised Ohm's law
\begin{equation}
\mathbf{E} = \frac{\mathbf{J} \times \mathbf{B}_{0}}{en_{e}} + \eta_{\text{eff}}\mathbf{J},\quad\quad\eta_{\text{eff}} = \frac{m_{e}\left( \nu - i\omega \right)}{e^{2}n_{e}}
\end{equation}
admits Beltrami solutions \(\nabla \times \mathbf{B} = \beta\mathbf{B}\) with \(\beta\) satisfying (Appendix B)
\begin{equation}
\beta^{2} - \frac{k_{z}\omega_{\text{ce}}}{\widetilde{\omega}}\beta + \frac{\omega \,\omega_{\text{pe}}^{2}}{c^{2}\widetilde{\omega}} = 0,\quad\quad\widetilde{\omega} = \omega + i\nu
\end{equation}
whence
\begin{equation}
\beta_{\pm} = \frac{k_{w}}{2}\left\lbrack 1 \pm \sqrt{1 - \frac{4k_{s}^{2}}{k_{w}^{2}}} \right\rbrack,\quad\quad k_{w} \equiv \frac{k_{z}\omega_{\text{ce}}}{\widetilde{\omega}},\quad\quad k_{s}^{2} \equiv \frac{\omega \,\omega_{\text{pe}}^{2}}{c^{2}\widetilde{\omega}}
\end{equation}
For \(4k_{s}^{2} \ll k_{w}^{2}\) the two classical limits are recovered:
\begin{equation}
\beta_{-} \simeq \frac{k_{s}^{2}}{k_{w}} = \frac{\omega\mu_{0}n_{e}e}{k_{z}B_{0}}\quad \Rightarrow \quad\beta k_{z} = \frac{\omega\mu_{0}n_{e}e}{B_{0}}\text{\quad\quad}\text{(helicon)}
\end{equation}
\begin{equation}
\beta_{+} \simeq k_{w} = \frac{k_{z}\omega_{\text{ce}}}{\omega}\text{\quad\quad}(\textrm{TG})
\end{equation}
The TG root exceeds the helicon root by the factor \(\Lambda\), a twenty-fold shorter wavelength for the reference case. Equivalently \(k_{\bot,\text{TG}}^{2} \simeq - (P/S)k_{z}^{2}\), which is the electrostatic resonance-cone condition \(\text{tan}^{2}\theta_{\text{res}} = - P/S\): the TG wave is the bounded-plasma manifestation of the resonance cone. This reduction is excellent in the developed discharge, i.e. at \(n_{e} = 10^{17}\) m$^{-3}$ its roots agree with the full quartic to \(2 \times 10^{- 3}\) relative error (Section 8.4), but it fails completely at ignition. Taking \(n_{e} \rightarrow 0\) gives \(k_{s} \rightarrow 0\) and hence
\begin{equation}
\beta_{-} \rightarrow 0,\quad\quad\beta_{+} \rightarrow k_{w} = k_{z}\Lambda \neq 0
\end{equation}
neither of which is the correct vacuum result \(\beta^{2} \rightarrow k_{0}^{2}\). The reason is that discarding the displacement current removes precisely the term that survives when the plasma current vanishes. The EMHD formulation therefore cannot be used to initialise a transient calculation; the full quartic, with displacement current and ion inertia retained, is required for \(n_{e} \lesssim n_{3}\).

\subsection{Coalescence and the critical density}
The discriminant of the EMHD quadratic vanishes when \(4k_{s}^{2} = k_{w}^{2}\). Collisionlessly this reads \(4\omega^{2}\omega_{\text{pe}}^{2}/c^{2} = k_{z}^{2}\omega_{\text{ce}}^{2}\), giving
\begin{equation}
n_{c} = \frac{k_{z}^{2}B_{0}^{2}}{4\mu_{0}m_{e}\omega^{2}}
\end{equation}
For the reference case \(n_{c} = 2.97 \times 10^{18}\) m$^{-3}$, squarely inside the operating range of a laboratory helicon source. Above \(n_{c}\) the two roots form a complex-conjugate pair: the helicon and TG waves cease to be separate normal modes and merge into a single coupled object with intrinsic radial decay. This is the coalescence identified by Shamrai and Taranov \cite{r12}, and it appears in the left panel of Figure 1 as an avoided crossing at which the two branches approach, exchange character and separate again.

Three consequences deserve emphasis. Since \(n_{c} \propto B_{0}^{2}\), the same source at 1 kG has \(n_{c} = 3 \times 10^{20}\) m$^{-3}$, far above any achievable density, so that the branches never coalesce and the classical well-separated picture holds throughout; at 100 G they coalesce within the operating range. This is the analytical content of the long-standing observation that TG physics matters at low magnetic field, and it is displayed as a regime map in Figure 2. Since \(n_{c} \propto k_{z}^{2}\), long axial wavelengths coalesce at lower density, so the several spectral components excited by a real antenna cross \(n_{c}\) at different times during a single ignition event. Finally, \(d\beta_{\pm}/dn_{e}\) diverges as \(\left( n_{c} - n_{e} \right)^{- 1/2}\) near coalescence, so adiabatic branch following must fail in a neighbourhood of \(n_{c}\) however slowly the density rises; this is treated in Section 5.5.
\begin{figure}[htbp]
\centering
\includegraphics[width=0.65\linewidth]{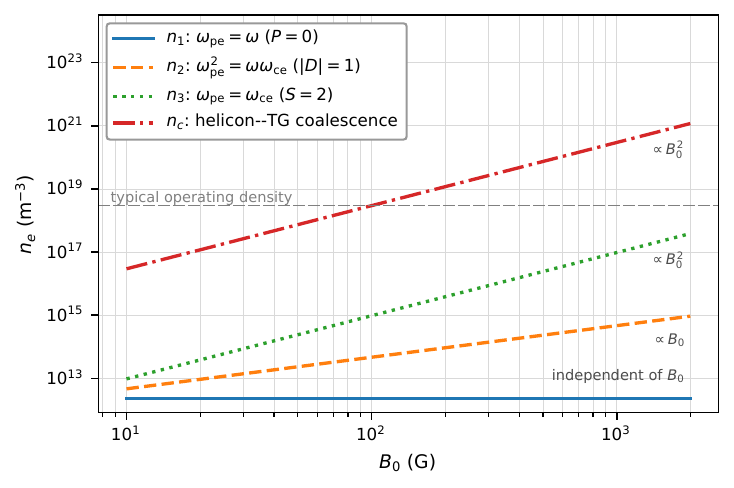}
\caption{Regime map in the \((B_{0},n_{e})\) plane at 13.56 MHz in argon. The three medium thresholds are spaced by powers of \(\Lambda = \omega_{\text{ce}}/\omega\) and fan out with field, while the coalescence density \(n_{c} \propto B_{0}^{2}\) rises much faster. Their intersection with the typical operating density (dotted) shows why TG physics dominates at low field and is irrelevant at high field.}
\label{fig:2}
\end{figure}

\subsection{Existence criterion for a lower-hybrid layer}
The coefficient \(\mathcal{A} = S\) vanishes at the lower-hybrid resonance, where the quartic degenerates and one root diverges, i.e. a genuine mode-conversion singularity \cite{r32}. Since
\begin{equation}
\frac{1}{\omega_{\text{LH}}^{2}} = \frac{1}{\omega_{\text{ci}}^{2} + \omega_{\text{pi}}^{2}} + \frac{1}{\omega_{\text{ce}}\omega_{\text{ci}}}
\end{equation}
\(\omega_{\text{LH}}\) rises from zero as \(n_{e}^{1/2}\) and saturates at \(\left( \omega_{\text{ce}}\omega_{\text{ci}} \right)^{1/2}\). A lower-hybrid layer therefore sweeps through the plasma during ignition if and only if
\begin{equation}
\omega < \sqrt{\omega_{\text{ce}}\,\omega_{\text{ci}}} = \omega_{\text{ce}}\sqrt{m_{e}/m_{i}}
\end{equation}
For argon at 100 G, \(\left( \omega_{\text{ce}}\omega_{\text{ci}} \right)^{1/2}/2\pi = 1.04\) MHz, far below 13.56 MHz: no lower-hybrid layer forms, \(S\) never vanishes and the quartic never degenerates. The criterion is not always satisfied, however. The same source at 1 kG gives 10.4 MHz, close to the drive frequency, and at 1.3 kG the layer appears; low-frequency sources in heavy gases cross it routinely. Where it does occur, a moving \(S = 0\) surface propagates inward from the edge carrying strong helicon$\rightarrow$TG conversion and a localised absorption spike, and the numerical treatment must resolve a moving singular layer.

\section{Bounded eigenvalue problem}
The local dispersion relation determines which wavenumbers the medium supports; it does not determine which the column supports. The transition to helicon operation is a transition of the bounded problem, and it is the boundary condition that converts a continuum of local roots into a discrete ladder of resonances.
\subsection{Field relations for a Beltrami branch}
For \(\nabla \times \mathbf{B} = \beta\mathbf{B}\), the radial and azimuthal components of the curl give
\begin{equation}
\frac{\text{im}}{r}B_{z} - ik_{z}B_{\theta} = \beta B_{r},\quad\quad ik_{z}B_{r} - \frac{\partial B_{z}}{\partial r} = \beta B_{\theta}
\end{equation}
and eliminating \(B_{\theta}\) yields the relation used throughout,
\begin{equation}
B_{r} = \frac{i}{T^{2}}\left( \frac{m\beta }{r}B_{z} + k_{z}\frac{\partial B_{z}}{\partial r} \right),\quad\quad T^{2} \equiv \beta^{2} - k_{z}^{2}
\end{equation}
Here \(B_{z}\) obeys Bessel's equation of order \(m\), so the regular solution is \(B_{z} = cJ_{m}(Tr)\) and the full field is the superposition of the two branches,
\begin{equation}
B_{z}(r) = c_{+}J_{m}\left( T_{+}r \right) + c_{-}J_{m}\left( T_{-}r \right),\quad\quad T_{\pm}^{2} = \beta_{\pm}^{2} - k_{z}^{2}
\end{equation}

\subsection{Boundary conditions and the dispersion function}
For a column bounded by an insulating tube with an external antenna, the appropriate pair of conditions is the vanishing of the radial current at the wall, \(J_{r}(a) = 0\), which for Beltrami fields reads \(\sum_{\pm}^{}\beta_{\pm}B_{r, \pm}(a) = 0\), together with the antenna drive \(B_{z}(a) = B_{a}\):
\begin{equation}
\sum_{j = \pm}^{}\frac{i\beta_{j}}{T_{j}^{2}}\left\lbrack \frac{m\beta_{j}}{a}J_{m}\left( T_{j}a \right) + k_{z}T_{j}J_{m}'\left( T_{j}a \right) \right\rbrack c_{j} = 0,\quad\quad\sum_{j = \pm}^{}J_{m}\left( T_{j}a \right)c_{j} = B_{a}
\end{equation}
This system determines the two amplitudes and is what is solved to produce Figure 7. Its homogeneous version has non-trivial solutions only when the determinant vanishes; that determinant is the dispersion function of the bounded problem, \(\mathcal{D}\left( \beta_{+},\beta_{-};m,k_{z},a \right) = 0\), whose roots in \(n_{e}\) at fixed \(\omega\) and \(k_{z}\) are the resonant densities. When the TG root is so strongly damped that it does not participate in the global resonance, the condition collapses to
\begin{equation}
\frac{m\beta }{a}J_{m}\left( \text{Ta} \right) + k_{z}TJ_{m}'\left( \text{Ta} \right) = 0
\end{equation}
which for \(m = 0\) gives \(J_{0}'(Ta) = 0\) and for \(m = 1\) with \(k_{z} \ll T\) gives \(J_{1}(Ta) = 0\), in both cases \(Ta = 3.832\).

\subsection{Resonance ladder and the resonance--coalescence ratio}
Combining \(T = 3.832/a\) with the helicon relation and \(\beta^{2} = k_{z}^{2} + T^{2}\) gives the resonant density of each axial mode \(p\) of a source of length \(L\),
\begin{equation}
n_{\text{res}}(p) = \frac{B_{0}k_{z}\sqrt{k_{z}^{2} + T^{2}}}{\omega\mu_{0}e},\quad\quad k_{z} = \frac{p\pi }{L},\quad\quad T = \frac{3.832}{a}
\end{equation}
This is the ladder the discharge climbs during ignition. At each rung the driven field becomes resonant with a bounded eigenmode, the antenna loading peaks and the absorbed power jumps, i.e. the microscopic origin of the E--H--W transitions and of the multiple jumps reported experimentally \cite{r24,r25,r33,r34}. Comparison with the coalescence density yields a compact and, to the authors' knowledge, previously unreported ratio:
\begin{equation}
\mathcal{R} \equiv \frac{n_{\text{res}}}{n_{c}} = \frac{B_{0}\beta k_{z}}{\omega\mu_{0}e} \cdot \frac{4\mu_{0}m_{e}\omega^{2}}{k_{z}^{2}B_{0}^{2}} = \frac{4\omega \beta }{\omega_{\text{ce}}k_{z}} = \frac{4}{\Lambda}\sqrt{1 + \frac{T^{2}}{k_{z}^{2}}}
\end{equation}
Setting \(\mathcal{R} = 1\) defines a critical axial wavenumber separating two qualitatively different classes of bounded mode,
\begin{equation}
k_{z}^{*} = \frac{T}{\sqrt{\Lambda^{2}/16 - 1}},\quad\quad\text{which\ exists\ only\ if\ }\Lambda > 4,\text{\ i.e.\ }\omega_{\text{ce}} > 4\omega .
\end{equation}
Modes with \(k_{z} < k_{z}^{*}\) have \(\mathcal{R} > 1\) and reach coalescence before their resonance, so that the resonance is embedded in the coalesced regime and is broad, lossy and poorly defined; modes with \(k_{z} > k_{z}^{*}\) resonate as clean, separated helicon modes and coalesce only later. If \(\omega_{\text{ce}} < 4\omega\), i.e. that is \(B_{0}\) below about 19 G at 13.56 MHz, no bounded mode resonates before coalescence at all. This is a sharp analytical lower bound on the field required for clean helicon operation, independent of density, power and gas.

For the reference case \(\Lambda = 20.64\) and \(T = 153.3\) m$^{-1}$ give \(k_{z}^{*} = 30.27\) m$^{-1}$ and \(p^{*} = k_{z}^{*}L/\pi = 3.85\). Table 3 and Figure 3 confirm that \(\mathcal{R}\) crosses unity between \(p = 3\) and \(p = 4\). The practical reading is that the low axial modes, i.e. those an antenna of finite length excites most strongly, are precisely the ones whose resonances are compromised by coalescence at 100 G, which argues for either raising \(B_{0}\) or shortening the antenna.
\begin{table}[htbp]
\centering
\footnotesize
\setlength{\tabcolsep}{3pt}
\caption{Bounded-mode resonance ladder compared with the coalescence density for each axial mode p. The ratio R crosses unity between p = 3 and p = 4.}
\label{tab:3}
\begin{tabularx}{\textwidth}{>{\raggedright\arraybackslash}X>{\raggedright\arraybackslash}X>{\raggedright\arraybackslash}X>{\raggedright\arraybackslash}X>{\raggedright\arraybackslash}X>{\raggedright\arraybackslash}X}
\toprule
\(p\) & \(k_{z}\) (m$^{-1}$) & \(\beta\) (m$^{-1}$) & \(n_{\text{res}}\) (m$^{-3}$) & \(n_{c}\) (m$^{-3}$) & \(\mathcal{R}\) \\
\midrule
1 & 7.85 & 153.5 & 7.03 $\times$ 10$^{17}$ & 1.86 $\times$ 10$^{17}$ & 3.79 \\
2 & 15.71 & 154.1 & 1.41 $\times$ 10$^{18}$ & 7.42 $\times$ 10$^{17}$ & 1.90 \\
3 & 23.56 & 155.1 & 2.13 $\times$ 10$^{18}$ & 1.67 $\times$ 10$^{18}$ & 1.28 \\
4 & 31.42 & 156.5 & 2.87 $\times$ 10$^{18}$ & 2.97 $\times$ 10$^{18}$ & 0.96 \\
5 & 39.27 & 158.2 & 3.62 $\times$ 10$^{18}$ & 4.64 $\times$ 10$^{18}$ & 0.78 \\
6 & 47.12 & 160.3 & 4.41 $\times$ 10$^{18}$ & 6.68 $\times$ 10$^{18}$ & 0.66 \\
\bottomrule
\end{tabularx}
\end{table}
\begin{figure}[htbp]
\centering
\includegraphics[width=0.7\linewidth]{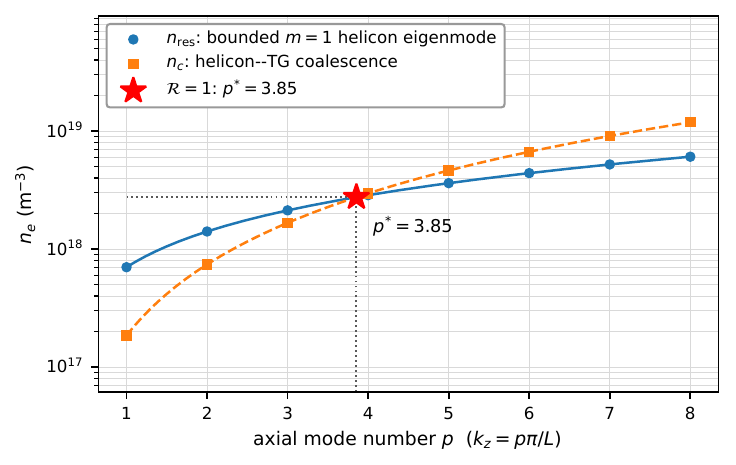}
\caption{The ladder of bounded-mode resonant densities \(n_{\text{res}}(p)\) climbed during ignition, compared with the coalescence density \(n_{c}(p)\) for the same axial mode. Markers are the physical integer modes; the curves continue \(p\) to real values so that the crossing is visible. The two curves cross between \(p = 3\) and \(p = 4\), at the analytically predicted \(k_{z}^{*} = T/(\Lambda^{2}/16 - 1)^{1/2}\). The star marks that crossing, \(\mathcal{R} = 1\) at \(p^{*} = 3.85\) and \(n_{e} = 2.76 \times 10^{18}\) m\(^{-3}\), with dotted guides to the axes.}
\label{fig:3}
\end{figure}

\subsection{Non-uniform profiles, turning points and the conversion layer}
Once \(n_{e} = n_{e}(r)\) the two-Bessel decomposition ceases to be exact and two complementary treatments are available. In the radial WKB description each branch takes the form, with $A_j$ a constant amplitude,
\begin{equation}
B_{z}(r) \simeq \frac{A_{j}}{\sqrt{r\,k_{\bot j}(r)}}\text{exp}\left\lbrack \pm i\int_{}^{r}k_{\bot j}(r')\, dr' \right\rbrack,\quad\quad\epsilon_{r} = \frac{1}{k_{\bot j}L_{n}} \ll 1
\end{equation}
For the TG branch \(k_{\bot} \sim 10^{2}\)--\(10^{3}\) m$^{-1}$ and \(L_{n} \sim a\) give \(\epsilon_{r} \sim 10^{- 2}\), so WKB is excellent; for the helicon branch \(k_{\bot} \sim T = 153\) m$^{-1}$ gives \(\epsilon_{r} \sim 0.3\) and WKB is marginal, so the global eigenvalue problem should be solved directly \cite{r35,r36,r37}. The appropriate hybrid strategy is therefore a full solve for the helicon branch combined with a WKB or boundary-layer treatment of the TG branch. The radial structure is organised by the zeros and poles of \(k_{\bot}^{2}(r)\). A zero, \(k_{\bot}^{2}(r) = 0\), is a turning point at which the wave reflects and beyond which the field decays; as \(n_{e}(r,t)\) rises the turning point moves. A pole at \(S(r) = 0\) is the lower-hybrid resonance, which by Section 3.8 does not occur for the reference case. The vanishing of the discriminant, \(n_{e}(r,t) = n_{c}\), defines the coalescence surface, i.e. the transient analogue of a mode-conversion layer, at which the two branches have equal wavenumber and exchange identity. Because \(n_{e}\) is centrally peaked in a developed discharge, the condition \(n_{e}(r_{c},t) = n_{c}\) is first met on axis, and the coalescence surface then expands outward as a growing cylindrical shell. Figure 4 shows this locus for a representative transport solution: the layer is born on axis at \(t \approx 320\) $\mu$s and sweeps to \(r/a \approx 0.37\) within 30 $\mu$s. Its speed
\begin{equation}
\frac{dr_{c}}{dt} = - \left. \ \frac{\partial n_{e}/\partial t}{\partial n_{e}/\partial r} \right|_{r_{c}}
\end{equation}
is set by the ratio of the ionisation rate to the profile steepness and can exceed the group velocity of the TG wave, in which case the conversion is non-adiabatic in space as well as in time.
\begin{figure}[htbp]
\centering
\includegraphics[width=0.95\linewidth]{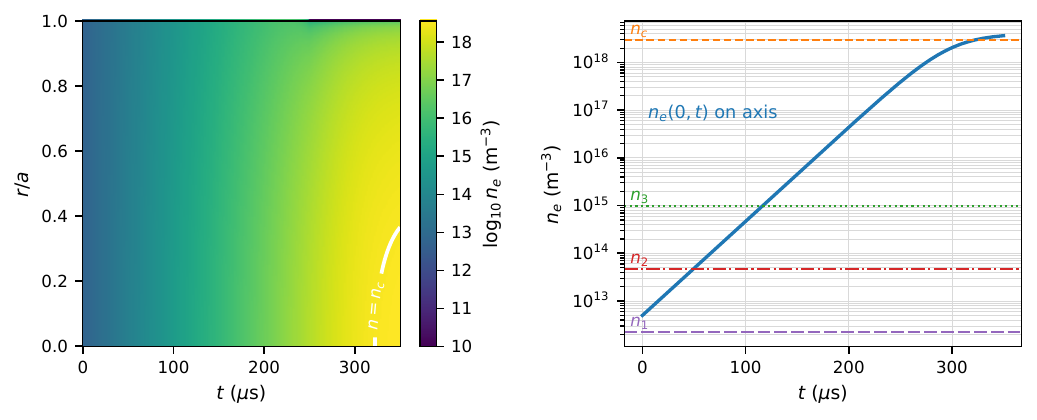}
\caption{Left: modelled density evolution \(n_{e}(r,t)\) with the profile shape morphing from flat to the diffusion eigenfunction, showing the locus \(n_{e}(r_{c},t) = n_{c}\) (white contour), the mode-conversion layer, which is born on axis and sweeps outward. Right: the on-axis density crossing the characteristic thresholds in the predicted order.}
\label{fig:4}
\end{figure}

\section{Two-timescale theory of the transient}
\subsection{Time-scale hierarchy}
Five time scales enter the problem, and their separation is what makes an analytical treatment possible:
\begin{equation}
\begin{gathered}
\omega^{- 1} \sim 10^{- 8}\, s\mspace{6mu} \ll \mspace{6mu}\nu_{m}^{- 1} \sim 10^{- 9}-10^{- 7}\, s\mspace{6mu} \lesssim \mspace{6mu}\tau_{\text{cav}} = \frac{2Q}{\omega} \sim 10^{- 7}-10^{- 5}\, s\mspace{6mu} \\
\ll \mspace{6mu}\tau_{n} = \left| \frac{{\dot{n}}_{e}}{n_{e}} \right|^{- 1} \sim 10^{- 5}-10^{- 4}\, s\mspace{6mu} \ll \mspace{6mu}\tau_{g} \sim 10^{- 3}\, s
\end{gathered}
\end{equation}
the last being the gas-depletion time. The interval between \(\omega^{- 1}\) and \(\tau_{n}\) spans four to five orders of magnitude and underwrites the adiabatic treatment of the constitutive law. The interval between \(\tau_{\text{cav}}\) and \(\tau_{n}\) spans at most two orders of magnitude and closes entirely near a resonance, where \(Q\) can exceed \(10^{3}\). That closure contains the whole of the non-adiabatic physics.

\subsection{Adiabatic branch tracking}
Writing \(\bar{t} = \xi t\) with \(\xi = \left( \omega\tau_{n} \right)^{- 1}\) and expanding \(\widetilde{\mathbf{E}} = {\widetilde{\mathbf{E}}}^{(0)} + \xi{\widetilde{\mathbf{E}}}^{(1)} + \cdots\), the leading order is the ordinary driven boundary-value problem with the medium frozen, in which $\mathbf{J_a}$ is the antenna current density,
\begin{equation}
\mathcal{L}\left\lbrack n_{e}(\bar{t}),\nu_{m}(\bar{t}) \right\rbrack{\widetilde{\mathbf{E}}}^{(0)} = - i\omega\mu_{0}\mathbf{J}_{a}
\end{equation}
The dispersion relation is then a one-parameter family \(\beta_{\pm} = \beta_{\pm}\left\lbrack n_{e}(\bar{t}) \right\rbrack\) and the transient is a trajectory through that family. This is the content of Figure 1, which read from left to right is the time history of the discharge.

\subsection{Non-adiabatic correction to the constitutive relation}
The leading order replaced \(\partial_{t} \rightarrow - i\omega\) in an equation whose coefficients are themselves time dependent, and the resulting error can be computed exactly. Defining
\begin{equation}
\mathcal{M} = \left( - i\omega + \nu_{\text{eff}} \right)\mathbb{I} + \mathbf{\Omega} \times ,\quad\quad\mathbf{\sigma} = \epsilon_{0}\omega_{\text{pe}}^{2}\mathcal{M}^{- 1}
\end{equation}
the exact equation is \(\mathcal{M}\widetilde{\mathbf{J}} + \xi\,\partial_{\bar{t}}\widetilde{\mathbf{J}} = \epsilon_{0}\omega_{\text{pe}}^{2}\widetilde{\mathbf{E}}\), and ordering gives
\begin{equation}
{\widetilde{\mathbf{J}}}^{(0)} = \mathbf{\sigma} \cdot {\widetilde{\mathbf{E}}}^{(0)},\quad\quad{\widetilde{\mathbf{J}}}^{(1)} = \mathbf{\sigma} \cdot {\widetilde{\mathbf{E}}}^{(1)} - \frac{1}{\epsilon_{0}\omega_{\text{pe}}^{2}}\,\mathbf{\sigma} \cdot \partial_{\bar{t}}\left( \mathbf{\sigma} \cdot {\widetilde{\mathbf{E}}}^{(0)} \right)
\end{equation}
The correction is a retardation term: the plasma current lags the instantaneous density. Its relative magnitude is
\begin{equation}
\frac{\left| \xi \,\partial_{\bar{t}}\widetilde{\mathbf{J}} \right|}{\left| \mathcal{M}\widetilde{\mathbf{J}} \right|} \sim \frac{\tau_{n}^{- 1}}{\sqrt{\omega^{2} + \omega_{\text{ce}}^{2}}} \simeq \frac{1}{\omega_{\text{ce}}\tau_{n}}\quad\text{(magnetised)},\quad\quad\epsilon_{\text{loc}} = \frac{1}{\omega\tau_{n}}\quad\text{(unmagnetised)}
\end{equation}
For the reference case with \(\tau_{n} \gtrsim 22\) $\mu$s, \(\epsilon_{\text{loc}} \simeq 5 \times 10^{- 4}\) and the magnetised estimate is \(2.6 \times 10^{- 5}\); Figure 5 confirms that \(\epsilon_{\text{loc}}\) never exceeds \(6 \times 10^{- 4}\) at any point in the transient. The local dielectric tensor may therefore be evaluated with the instantaneous density and collision frequency throughout ignition, with fractional error below \(10^{- 3}\); no memory kernel, convolution or time-domain constitutive equation is required. Equivalently, in the frequency domain the modulation of the medium generates sidebands of width \(\Delta\omega \sim \tau_{n}^{- 1} \sim 5 \times 10^{4}\) s$^{-1}$, a fractional smearing of the dispersion curve of order \(10^{- 4}\).
\begin{figure}[htbp]
\centering
\includegraphics[width=0.7\linewidth]{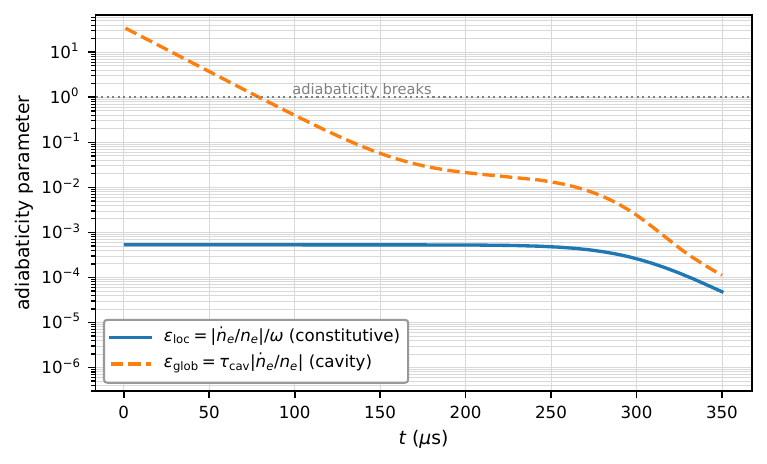}
\caption{The two adiabaticity parameters through the transient. The local parameter \(\epsilon_{\text{loc}} = |{\dot{n}}_{e}/n_{e}|/\omega\) never exceeds \(6 \times 10^{- 4}\), so the instantaneous dielectric tensor is valid throughout. The global parameter \(\epsilon_{\text{glob}} = \tau_{\text{cav}}|{\dot{n}}_{e}/n_{e}|\) exceeds unity for the first $\sim$80 \(\mu\)s, where the field cannot follow the density. All non-adiabatic physics of ignition lies in this gap.}
\label{fig:5}
\end{figure}

\subsection{Global non-adiabaticity and the swept resonance}
The bounded-mode amplitude behaves quite differently. Near an isolated resonance the driven amplitude \(A(t)\) of a normal mode obeys
\begin{equation}
\frac{dA}{dt} = \left\lbrack i\,\delta(t) - \gamma \right\rbrack A + F
\end{equation}
with \(F\) the antenna drive, \(\gamma = \omega/2Q\) the damping rate and \(\delta(t) = \omega - \omega_{\text{res}}\left\lbrack n_{e}(t) \right\rbrack\) the instantaneous detuning, swept through zero as the density climbs past \(n_{\text{res}}\). Linearising the sweep as \(\delta \simeq \alpha\left( t - t_{\text{res}} \right)\) with
\begin{equation}
\alpha = \left| \frac{\partial\omega_{\text{res}}}{\partial\text{ln}n_{e}} \right|\frac{1}{\tau_{n}}
\end{equation}
the control parameter is
\begin{equation}
\mathcal{S} = \frac{\alpha}{\gamma^{2}} = 4Q^{2}\,\frac{\left| d\omega_{\text{res}}/dt \right|}{\omega^{2}}
\end{equation}
and the solution is the Fresnel-type integral
\begin{equation}
A(t) = F\int_{- \infty}^{t}\text{exp}\left\lbrack - \gamma\left( t - t' \right) + \frac{i\alpha }{2}\left( t^{2} - t'^{2} \right) \right\rbrack dt'
\end{equation}

Figure 6 displays its behaviour. For \(\mathcal{S} \ll 1\) the amplitude traces the static Lorentzian, peaking at \(F/\gamma\) exactly at \(\delta = 0\). For \(\mathcal{S} \sim 1\) the peak is delayed past resonance by \(\Delta\delta \sim \gamma\mathcal{S}^{1/2}\), reduced in height, and followed by chirped oscillations of instantaneous frequency \(\alpha\left( t - t_{\text{res}} \right)\). For \(\mathcal{S} \gg 1\) the mode is swept through before it can fill; the peak amplitude scales as \(F\left( \pi/\alpha \right)^{1/2}\) rather than \(F/\gamma\), a reduction by \(\left( \pi/\mathcal{S} \right)^{1/2}\) up to a numerical factor: exact evaluation of the Fresnel integral gives \(\gamma\left| A \right|_{\max} \rightarrow 1.66\left( \pi/\mathcal{S} \right)^{1/2}\) as \(\mathcal{S} \rightarrow \infty\), the constant being the Cornu-spiral maximum. The peak occurs far downstream of the nominal resonance.

Figure 5 evaluates \(\epsilon_{\text{glob}} = \tau_{\text{cav}}\left| {\dot{n}}_{e}/n_{e} \right|\) for the reference transient. It exceeds unity for the first $\sim$80 $\mu$s and falls below \(10^{- 2}\) thereafter. The early transient is thus globally non-adiabatic and the late transient globally adiabatic. A quasi-static description is quantitatively wrong for the first tens of microseconds, i.e. it overestimates the field amplitude at every early resonance crossing and misses the delay, the ringing and the hysteresis, but becomes accurate once the discharge is established. The observable signatures are that the resonance appears at a higher density than \(n_{\text{res}}\) predicts, by a fraction \(\sim \mathcal{S}^{1/2}/2Q\); that the loading peak is lower and broader than the static calculation; and that each crossing is followed by a decaying oscillation whose frequency rises with time.
\begin{figure}[htbp]
\centering
\includegraphics[width=0.65\linewidth]{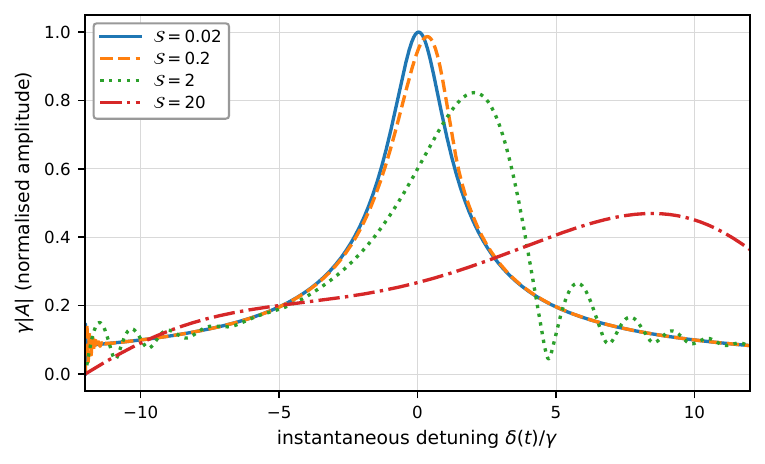}
\caption{Exact solution of the swept-resonance problem for four values of the sweep parameter \(\mathcal{S} = |d\delta/dt|/\gamma^{2}\). As \(\mathcal{S}\) grows the response peak is delayed past resonance, reduced in height in proportion to \(\mathcal{S}^{- 1/2}\), and followed by chirped ringing. A quasi-static calculation reproduces only the \(\mathcal{S} \rightarrow 0\) curve.}
\label{fig:6}
\end{figure}

\subsection{Landau--Zener conversion at coalescence}
Near \(n_{c}\) the two branches undergo an avoided crossing. Reducing to the two-state form with a slowly varying parameter,
\begin{equation}
i\frac{d}{dz}\begin{pmatrix}
a_{\text{Helicon}} \\
a_{\text{TG}} \\
\end{pmatrix} = \begin{pmatrix}
\beta_{\text{Helicon}}(z) & \kappa_{\text{LZ}} \\
\kappa_{\text{LZ}}^{*} & \beta_{\text{TG}}(z) \\
\end{pmatrix}\begin{pmatrix}
a_{\text{Helicon}} \\
a_{\text{TG}} \\
\end{pmatrix}
\end{equation}
in which \(\kappa_{\text{LZ}}\) is the off-diagonal coupling between the two branches, distinct from the vacuum evanescence rate \(\kappa\) of Section 3.2. The probability of remaining on the original branch \cite{r38,r39}, that is of failing to convert, is
\begin{equation}
P_{\text{na}} = \text{exp}\left( - 2\pi \Gamma  \right),\quad\quad\Gamma = \frac{\left| \kappa_{\text{LZ}} \right|^{2}}{\left| d\left( \beta_{\text{Helicon}} - \beta_{\text{TG}} \right)/dz \right|}
\end{equation}
The gap at closest approach is set by collisions. At exact coalescence the EMHD quadratic gives
\begin{equation}
1 - \frac{4k_{s}^{2}}{k_{w}^{2}} = - \,\frac{i\nu }{\omega}\quad\quad \Rightarrow \quad\quad\beta_{+} - \beta_{-} = k_{w}\sqrt{- \frac{i\nu }{\omega}},\quad\quad\left| \kappa_{\text{LZ}} \right| = \frac{1}{2}\left| \beta_{+} - \beta_{-} \right| \simeq \frac{k_{z}\Lambda}{2}\sqrt{\frac{\nu}{\omega}}
\end{equation}
so the coupling, and hence the conversion efficiency, scales as \(\left( \nu/\omega \right)^{1/2}\): a collisionless plasma converts poorly and a collisional one efficiently. This is the microscopic reason that TG damping is the dominant absorption channel in real, moderately collisional helicon sources. The sweep rate \(d\left( \beta_{\text{Helicon}} - \beta_{\text{TG}} \right)/dz\) is the spatial gradient produced by the radial density profile, so that \(\Gamma \propto \nu L_{n}\): conversion is efficient for collisional plasmas with gentle profiles and inefficient for collisionless plasmas with steep edges. During ignition \(\nu\) falls and \(L_{n}\) steepens, so \(\Gamma\) decreases monotonically and the discharge converts efficiently early and inefficiently late. This predicts a maximum in fractional edge absorption partway through the transient, which is what Figure 7 shows: 94\% at stage III, falling to 78\% at stage IV.

\subsection{Sudden limit and mode projection}
If the density changes faster than the cavity can respond, which the E--H and H--W jumps can themselves produce, i.e. being self-accelerating, the field does not follow and must be re-expanded in the new eigenbasis,
\begin{equation}
\widetilde{\mathbf{E}}\left( t_{+} \right) = \sum_{j}^{}c_{j}\mathbf{\psi}_{j}^{\text{new}},\quad\quad c_{j} = \frac{\left\langle \mathbf{\psi}_{j}^{\text{new}}|\widetilde{\mathbf{E}}\left( t_{-} \right) \right\rangle}{\left\langle \mathbf{\psi}_{j}^{\text{new}}|\mathbf{\psi}_{j}^{\text{new}} \right\rangle}
\end{equation}
with the inner product weighted by \(\partial\left( \omega\mathbf{\varepsilon} \right)/\partial\omega\) so that it is the wave-energy norm. Because the vacuum field \(I_{m}(\kappa r)\) is monotonic and edge-peaked while the developed helicon eigenmode \(J_{m}(Tr)\) is oscillatory with interior structure, the overlap is poor, typically \(\left| c_{\text{Helicon}} \right|^{2} \lesssim 0.3\) for \(m = 1\). A sudden transition therefore deposits most of the stored field energy into non-resonant, strongly damped components rather than into the helicon mode, which is a quantitative argument for slow power ramps where efficient wave coupling is the objective.

\subsection{Wave action rather than wave energy}
In a medium whose properties change in time, wave energy is not conserved even in the absence of dissipation, because the medium does work on the wave; the adiabatically conserved quantity \cite{r40} is the wave action \(\mathcal{N} = W/\omega\) with
\begin{equation}
W = \frac{\epsilon_{0}}{4}\,{\widetilde{\mathbf{E}}}^{*} \cdot \frac{\partial\left( \omega\mathbf{\varepsilon}_{H} \right)}{\partial\omega} \cdot \widetilde{\mathbf{E}} + \frac{1}{4\mu_{0}}\left| \widetilde{\mathbf{B}} \right|^{2}
\end{equation}
where \(\mathbf{\varepsilon}_{H}\) is the Hermitian part of the tensor. In the driven problem \(\omega\) is fixed by the antenna, so action and energy remain proportional and no frequency shift occurs; but the coefficient \(\partial\left( \omega\mathbf{\varepsilon} \right)/\partial\omega\) changes by orders of magnitude during the transient. A given field amplitude therefore corresponds to very different stored energies at different stages of the discharge, and comparisons between time steps should be made in terms of \(W\) or \(\mathcal{N}\) rather than \(\left| \mathbf{E} \right|\).

\section{Coupling to plasma formation and transport}
The wave theory above treats \(n_{e}(r,t)\) and \(\nu_{m}(r,t)\) as given. This section supplies them at the level required to close the problem and to fix the time scales. Global particle balance, in which $n_g$ is the neutral density, $K_{\text{iz}}$ the ionisation rate coefficient and $\tau_{\text{loss}}$ the particle confinement time,
\begin{equation}
\frac{dn_{e}}{dt} = n_{e}n_{g}K_{\text{iz}}\left( T_{e} \right) - \frac{n_{e}}{\tau_{\text{loss}}},\quad\quad\tau_{n}^{- 1} = n_{g}K_{\text{iz}} - \tau_{\text{loss}}^{- 1}
\end{equation}
gives exponential growth during the avalanche. For argon at 10--50 mTorr with \(T_{e} = 3\) eV, \(K_{\text{iz}} \sim 10^{- 16}\) m$^{3}$ s$^{-1}$ and \(n_{g} = 3 \times 10^{20}\)--\(1.6 \times 10^{21}\) m$^{-3}$ give \(n_{g}K_{\text{iz}} \sim 3 \times 10^{4}\)--\(1.6 \times 10^{5}\) s$^{-1}$, so that with wall losses at the Bohm rate \cite{r41,r42} \(\tau_{n}\) is typically 20--100 $\mu$s. Because \(K_{\text{iz}}\) is exponentially sensitive to \(T_{e}\), which is set by the absorbed power and hence by the wave field, the transient is intrinsically a feedback loop; it is this feedback that converts the resonance ladder of Section 4.3 into the sharp and sometimes hysteretic mode jumps observed experimentally \cite{r43,r44}.

Radially, \(\partial_{t}n_{e} = \nabla \cdot \left( D_{a}\nabla n_{e} \right) + S_{\text{iz}}\) with \(D_{a} \simeq D_{i}\left( 1 + T_{e}/T_{i} \right)\) across the field, where $D_a$ is the ambipolar diffusion coefficient, $D_i$ the ion diffusion coefficient and $T_i$ the ion temperature. The profile shape is set by the competition between the ionisation source, which follows the power deposition and is therefore edge-localised whenever the TG branch dominates, and diffusion, which relaxes towards the fundamental eigenfunction \(J_{0}\left( 2.405\, r/a \right)\). Three stages follow. Early, absorption is a collisional skin effect at the edge, the source is edge-peaked and diffusion has not yet acted, so the profile is flat to edge-peaked. In the middle stage TG absorption is strongly edge-localised, reinforcing an edge source, while the diffusion time \(a^{2}/D_{a}\) becomes comparable to \(\tau_{n}\) and begins to fill the centre. Late, the helicon branch deposits power volumetrically, the source moves inward and the profile becomes centrally peaked. The shape feeds back on the wave through \(T\) and through the location of the coalescence surface, closing the loop.

Two effects reduce \(\nu_{m}\) during ignition, both fast enough to matter. Gas heating at constant pressure gives \(n_{g} \propto 1/T_{g}\), and \(T_{g}\) rises from 300 K to 600--1000 K in a few hundred microseconds \cite{r45}, reducing \(\nu_{m}\) by a factor of 2--3. Neutral pump-out by ionisation and ion--neutral drag produces on-axis depletion factors of 3--10 in helicon sources \cite{r46,r47,r48}. Together these can drop \(\nu_{m}\) by an order of magnitude, which is precisely the change required to cross the demagnetisation threshold of Section 3.5. Neutral dynamics are therefore not a refinement but part of the mechanism by which the discharge becomes a helicon discharge, and a model holding \(n_{g}\) fixed will not reproduce the correct E--H--W timing.

All of this is invisible from outside the tube except through the plasma loading resistance seen by the matching network,
\begin{equation}
R_{p}\left( n_{e} \right) = \frac{2P_{\text{abs}}}{\left| I_{a} \right|^{2}}
\end{equation}
which peaks at each rung of the resonance ladder. The measured \(R_{p}(t)\) during ignition is therefore the natural experimental trace against which to test the theory, and is far more robust than any internal field measurement.

\section{Power deposition}
The cycle-averaged absorbed power density is
\begin{equation}
p(\mathbf{r}) = \frac{1}{2}\,\text{Re}\left( \mathbf{J}^{*} \cdot \mathbf{E} \right) = \frac{\omega\epsilon_{0}}{2}\,{\widetilde{\mathbf{E}}}^{*} \cdot \mathbf{\varepsilon}_{A} \cdot \widetilde{\mathbf{E}}
\end{equation}
with \(\mathbf{\varepsilon}_{A}\) the anti-Hermitian part of the tensor. In the EMHD limit this simplifies usefully: since \(\left( \mathbf{J} \times \mathbf{B}_{0} \right) \cdot \mathbf{J}^{*} = \mathbf{B}_{0} \cdot \left( \mathbf{J}^{*} \times \mathbf{J} \right)\) is purely imaginary for any complex \(\mathbf{J}\), the magnetic term does no net work and
\begin{equation}
p(r) = \frac{1}{2}\,\text{Re}\left( \eta_{\text{eff}} \right)\left| \mathbf{J}(r) \right|^{2} = \frac{m_{e}\nu_{\text{eff}}}{2e^{2}n_{e}}\left| \mathbf{J}(r) \right|^{2}
\end{equation}
Each branch carries \(\mathbf{J}_{j} = \beta_{j}\mathbf{B}_{j}/\mu_{0}\), so the TG branch, whose \(\left| \beta_{+} \right|\) is larger by \(\Lambda\), carries \(\Lambda^{2} \approx 426\) times more current per unit field. Combined with its short damping length \(\delta_{\text{TG}} = 1/\text{Im}\,\beta_{+}\), this makes the TG branch the dominant absorber over most of the transient even when it carries a small fraction of the field. Table 4 gives the computed values; the corresponding measured and modelled deposition profiles for a blue-core discharge are given in \cite{r49}. Reading down the table, \(\lambda_{\text{TG}}\) collapses from 12 m, i.e. larger than the entire device so that the TG ``wave'' is a global evanescent structure rather than a wave, to 16 mm, comparable to the tube radius. The skin depth falls to 3 mm at the point of maximum edge localisation, and \(Q_{\text{Helicon}}\) falls from $\sim$1700 to below unity as the helicon branch becomes lossy past coalescence; the sign reversal of \(\text{Im}\,\beta \) in the last row marks the post-coalescence regime. The quality factor \(Q_{\text{Helicon}}\) is defined as \(\text{Re}\,\beta/2\left| \text{Im}\,\beta \right|\) for the helicon branch. Figure 7 shows the corresponding eigenfunctions and deposition profiles.
\begin{table}[htbp]
\centering
\footnotesize
\setlength{\tabcolsep}{3pt}
\caption{Branch wavenumbers, TG wavelength and skin depth, and helicon quality factor through the ignition transient.}
\label{tab:4}
\begin{tabularx}{\textwidth}{>{\raggedright\arraybackslash}X>{\raggedright\arraybackslash}X>{\raggedright\arraybackslash}X>{\raggedright\arraybackslash}X>{\raggedright\arraybackslash}X>{\raggedright\arraybackslash}X>{\raggedright\arraybackslash}X}
\toprule
\(n_{e}\) (m$^{-3}$) & \(\nu\) (s$^{-1}$) & \(\beta_{\text{Helicon}}\) (m$^{-1}$) & \(\beta_{\text{TG}}\) (m$^{-1}$) & \(\lambda_{\text{TG}}\) (mm) & \(\delta_{\text{TG}}\) (mm) & \(Q_{\text{Helicon}}\) \\
\midrule
10$^{14}$ & 3 $\times$ 10$^{9}$ & 0.0 + 0.00\(i\) & 0.5 - 18.4\(i\) & 12149 & 54.3 & 1687 \\
10$^{15}$ & 3 $\times$ 10$^{9}$ & 0.1 + 0.00\(i\) & 0.5 - 18.4\(i\) & 13424 & 54.3 & 169 \\
10$^{16}$ & 1 $\times$ 10$^{9}$ & 0.5 + 0.01\(i\) & 4.1 - 54.9\(i\) & 1522 & 18.2 & 50 \\
10$^{17}$ & 3 $\times$ 10$^{8}$ & 5.5 + 0.17\(i\) & 42.9 - 170.6\(i\) & 146 & 5.86 & 17 \\
3 $\times$ 10$^{17}$ & 1 $\times$ 10$^{8}$ & 16.8 + 0.54\(i\) & 256 - 321\(i\) & 24.5 & 3.12 & 16 \\
10$^{18}$ & 3 $\times$ 10$^{7}$ & 60.0 + 2.39\(i\) & 517 - 206\(i\) & 12.2 & 4.87 & 13 \\
3 $\times$ 10$^{18}$ & 1 $\times$ 10$^{7}$ & 255 + 52.6\(i\) & 385 - 128\(i\) & 16.3 & 7.84 & 2.4 \\
10$^{19}$ & 1 $\times$ 10$^{7}$ & 303 - 536\(i\) & 337 + 461\(i\) & 18.7 & 2.17 & 0.3 \\
\bottomrule
\end{tabularx}
\end{table}
\begin{figure}[htbp]
\centering
\includegraphics[width=0.95\linewidth]{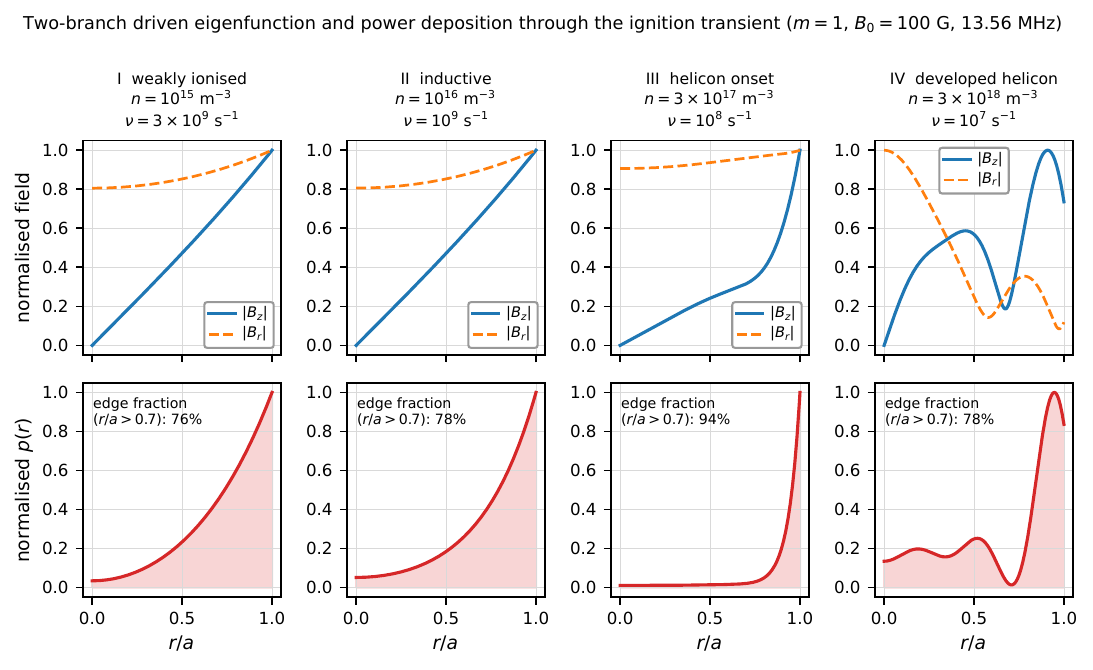}
\caption{Two-branch driven eigenfunction (top) and power deposition profile (bottom) at four stages of ignition, from the boundary-value problem of Section 4.2. The vacuum-like \(|B_{z}| \propto r\) near field of stage I develops interior structure only once the helicon branch propagates. Stage IV lies at \(n_{e}/n_{c} = 1.01\), that is essentially at coalescence and at the \(p = 4\) bounded resonance, which is why its profile shows both interior structure and a residual edge peak. Edge localisation of the absorbed power reaches 94\% in stage III, the TG-dominated phase, and falls to 78\% thereafter, as predicted by the scaling \(\Gamma \propto \nu L_{n}\) of Section 5.5.}
\label{fig:7}
\end{figure}

The cold-plasma treatment omits Landau damping, which for the TG branch becomes comparable to collisional damping once \(\omega/k_{z} \sim v_{\text{te}}\). For the reference case \(\omega/k_{z} = 2.7 \times 10^{6}\) m s$^{-1}$ against \(v_{\text{te}} = 7.3 \times 10^{5}\) m s$^{-1}$ at 3 eV, so \(\omega/k_{z}v_{\text{te}} \approx 3.7\): Landau damping is present but not dominant. Here $v_{\text{te}}$ is the electron thermal speed $(T_e/m_e)^{1/2}$. It matters specifically for the TG branch, whose large \(\beta\) reduces the parallel phase velocity of the converted wave. A cold model will therefore underestimate edge absorption in the late transient by a factor that grows as the plasma becomes less collisional. This is the principal limitation of the theory presented here and the natural direction for its extension.

\section{Implications for numerical modelling and experiment}
The theory is analytical, but every result carries a consequence for the numerical treatment of the same problem. This section states those consequences in a form independent of any particular code, together with benchmark values and testable predictions.
\subsection{Formulation requirements}
Two requirements follow directly from Sections 2.4 and 3.6. The wave problem must be posed in terms of the relative permittivity tensor rather than a conductivity, because only the former reduces to the identity as \(n_{e} \rightarrow 0\); a conductivity formulation divides by \(n_{e}\) and becomes ill-conditioned at ignition. Moreover, the EMHD or Beltrami reduction, although convenient in the developed discharge, cannot be used to initialise a transient calculation, since its roots do not approach the vacuum values. Ion terms in the tensor cost nothing and guarantee correct behaviour should the operating point approach the lower-hybrid condition of Section 3.8.

\subsection{Resolution set by the TG layer}
The binding constraint on spatial resolution is the TG branch, not the helicon branch. From Table 4 the most demanding case is \(n_{e} \approx 3 \times 10^{17}\) m$^{-3}$, where \(\lambda_{\text{TG}} = 24.5\) mm but \(\delta_{\text{TG}} = 3.1\) mm. Requiring eight elements per wavelength and four across the skin depth gives
\begin{equation}
h < \text{min}\left( \frac{\lambda_{\text{TG}}}{8},\mspace{6mu}\frac{\delta_{\text{TG}}}{4} \right) = 0.78\ \text{mm}
\end{equation}
about 32 elements across the 25 mm radius as an absolute minimum. Since the layer sits at the edge, a graded boundary-layer mesh is far more efficient than uniform refinement. Critically, \(\delta_{\text{TG}}\) varies by a factor of 25 across the transient, from 54 mm to 2.2 mm, so a single fixed mesh will be either wasteful or inadequate; the mesh should be keyed to \(\delta_{\text{TG}}\left( n_{e} \right)\) and re-evaluated as the density evolves. A useful convergence diagnostic is the number of elements per skin depth evaluated from \(\text{Im}\,\beta_{+}\) analytically; where it falls below four, the absorbed power is not converged even if the field profile appears smooth.

\subsection{A staged verification strategy}
Sections 5.2--5.4 imply a specific and economical sequence of numerical experiments. A frequency-domain parametric sweep in \(n_{e}\) at fixed profile shape computes the adiabatic family \(\beta_{\pm}\left( n_{e} \right)\) directly and is the cheapest informative experiment; \(\nu_{m}\) must be swept alongside \(n_{e}\) rather than held constant, or the demagnetisation threshold is never crossed. Repeating the sweep with flat, parabolic and \(J_{0}\) profiles isolates the effect of profile shape from that of magnitude and locates the coalescence surface. A time-dependent solve with prescribed \(n_{e}(r,t)\) then isolates the global non-adiabatic physics, and the comparison between the two is itself the diagnostic: the quasi-static and transient results should agree wherever \(\epsilon_{\text{glob}} < 1\) and diverge wherever it exceeds unity. Only after these should the fully coupled wave--transport problem be attempted, as in \cite{r27,r50}, since the earlier stages establish what the answer ought to look like.

\subsection{Benchmark values and verification of the analytical reductions}
Table 5 lists six benchmarks with the values the model must reproduce, and Table 7 of Appendix C collects the derived quantities they are checked against. The first is the one most often omitted and most often the source of later trouble: a model that does not reproduce the vacuum modified-Bessel near field to 1\% is not trustworthy downstream.
\begin{table}[htbp]
\centering
\footnotesize
\setlength{\tabcolsep}{3pt}
\caption{Benchmark values for verification of a numerical model, in the order the tests should be performed.}
\label{tab:5}
\begin{tabularx}{\textwidth}{l>{\raggedright\arraybackslash\hsize=0.747\hsize}X>{\raggedright\arraybackslash\hsize=1.253\hsize}Xl}
\toprule
Test & Configuration & Expected result & Tolerance \\
\midrule
B1 & \(n_{e} = 10^{6}\) m$^{-3}$ (numerical vacuum) & \(\left| B_{z} \right| \propto I_{1}(\kappa r)\), \(\kappa = 31.4\) m$^{-1}$; no absorption & 1\% on \(\kappa\) \\
B2 & \(n_{e} = 10^{17}\) m$^{-3}$, \(\nu = 3 \times 10^{8}\) s$^{-1}$ & \(\beta_{\text{Helicon}} = 5.5 + 0.17i\), \(\beta_{\text{TG}} = 42.9 - 170.6i\) m$^{-1}$ & 2\% \\
B3 & \(n_{e} = 10^{17}\) m$^{-3}$, \(\nu \rightarrow 0\) & \(\beta_{\text{Helicon}}k_{z} = \omega\mu_{0}n_{e}e/B_{0} = 173\) m$^{-2}$ & 1\% \\
B4 & Sweep \(n_{e}\), locate loading peak & \(n_{\text{res}} = 2.87 \times 10^{18}\) m$^{-3}$ for \(p = 4\) & 5\% \\
B5 & Sweep \(n_{e}\) through coalescence & Branch exchange at \(n_{c} = 2.97 \times 10^{18}\) m$^{-3}$ & 5\% \\
B6 & Transient with \(\tau_{n} = 22\) $\mu$s & Loading peak delayed relative to B4; chirped ringing & qualitative \\
\bottomrule
\end{tabularx}
\end{table}
The closed-form reductions used throughout were themselves verified against direct numerical evaluation of the full quartic. Table 6 reports the residuals. Those of order \(10^{- 2}\) are the expected size of the neglected terms in each asymptotic limit, given \(\omega/\omega_{\text{ce}} = 0.048\), which confirms that the reductions are correct rather than merely self-consistent.
\begin{table}[htbp]
\centering
\footnotesize
\setlength{\tabcolsep}{3pt}
\caption{Residuals of the closed-form reductions against direct numerical evaluation of the full quartic.}
\label{tab:6}
\begin{tabularx}{\textwidth}{>{\raggedright\arraybackslash}X>{\raggedright\arraybackslash}X}
\toprule
Verified relation & Relative residual \\
\midrule
Vacuum limit is an exact double root & 0.0 \\
Helicon relation \(\beta k_{z} = \omega\mu_{0}n_{e}e/B_{0}\) & 8.6 $\times$ 10$^{-3}$ \\
TG limit \(\beta_{+} = k_{z}\omega_{\text{ce}}/\omega\) & 8.5 $\times$ 10$^{-3}$ \\
EMHD reduction against full quartic at \(10^{17}\) m$^{-3}$ & 2.0 $\times$ 10$^{-3}$ \\
Coalescence density, analytical against numerical root & 5.4 $\times$ 10$^{-10}$ \\
TG perpendicular wavenumber against \(\sqrt{- P/S}\, k_{z}\) & 8.5 $\times$ 10$^{-3}$ \\
\bottomrule
\end{tabularx}
\end{table}

\subsection{Testable predictions}
The theory makes following predictions for numerical and experimental verifications.
\begin{enumerate}
\item As \(n_{e} \rightarrow 0\) the two computed roots converge to the single value \(k_{\bot}^{2} = k_{0}^{2} - k_{z}^{2}\), and the splitting grows linearly in \(n_{e}\), by a factor of ten for every tenfold rise in density. The linear law is asymptotic and should be tested over \(10^{8} \lesssim n_{e} \lesssim 10^{12}\) m$^{-3}$, where it holds to better than 1\%; by \(n_{e} = 10^{14}\) m$^{-3}$ the decade ratio has fallen to 9.2 as \(n_{3}\) is approached and the expansion in \(D\) ceases to be small.
\item The upper root crosses \(\left| \beta \right| = k_{z}\) at \(n_{1}\) and the lower root at \(n_{\text{cut}} = (4/\Lambda)n_{c}\), the latter remaining within 1\% of \(k_{0}\) until \(n_{e} \simeq n_{3}\).
\item If \(\nu_{m}\) is held above \(\omega_{\text{ce}}\), no helicon--TG separation occurs at any density.
\item The coalescence point scales as \(B_{0}^{2}k_{z}^{2}\); repeating a sweep at 300 G moves the branch exchange up by a factor of nine.
\item \(\mathcal{R}\) crosses unity between \(p = 3\) and \(p = 4\) at 100 G, and no bounded mode resonates before coalescence if \(B_{0} < 19\) G.
\item A time-domain constitutive model and an instantaneous-tensor model differ by less than \(10^{- 3}\).
\item Quasi-static and transient calculations diverge exactly where \(\epsilon_{\text{glob}} > 1\), that is during the first $\sim$80 $\mu$s.
\item The transient loading peak occurs at higher density than the static prediction and is followed by a decaying oscillation of rising frequency.
\item The fraction of power deposited outside \(r/a = 0.7\) peaks near \(n_{e} \sim 3 \times 10^{17}\) m$^{-3}$ at about 94\% and falls thereafter.
\item A slow power ramp couples more energy into the helicon mode than a step of equal total energy, by approximately the overlap-integral factor of Section 5.6.
\end{enumerate}
Predictions 3, 5, 8 and 10 are accessible to experiment through the loading resistance \(R_{p}(t)\) alone; the remainder require either internal field measurements or numerical solution.

\section{Conclusions}
This paper has developed an analytical theory of how the wave dispersion relation of a helicon discharge evolves from the vacuum state at breakdown to the steady, wave-coupled state, using a cold, collisional, two-species description that remains regular in the vacuum limit and therefore contains both end states within a single formalism.

The central structural result is that the vacuum limit of the quartic governing the perpendicular wavenumber is a degenerate double root. The helicon and TG branches are consequently not independent modes that appear at some threshold density but the two lobes of a single vacuum root, separated by the gyrotropic element \(D\) and the anisotropy \(P - S\) of the dielectric tensor. The degeneracy is broken more weakly than the general theory of perturbed repeated roots implies: the first-order term of the discriminant cancels identically, so the branches separate linearly in \(n_{e}\) rather than as \(n_{e}^{1/2}\). The separation is therefore suppressed, and the two roots emerge on very different schedules. The TG root departs from the vacuum value immediately and becomes radially propagating exactly at \(n_{1} = \epsilon_{0}m_{e}\omega^{2}/e^{2}\), where \(P\) changes sign; the helicon root remains pinned at \(\beta = k_{0}\) to within 1\% until \(n_{e} \simeq n_{3}\) and does not propagate until \(n_{\text{cut}} = (4/\Lambda)n_{c}\). Some five orders of magnitude in density therefore separate the establishment of the TG branch from the existence of the helicon branch at all, which is the analytical statement of why a helicon source must pass through a long inductive phase before it can couple as a wave.

The transient is organised by five characteristic densities and one magnetisation condition. The three medium thresholds \(n_{1}\), \(n_{2}\), \(n_{3}\) stand in the ratio \(1:\Lambda:\Lambda^{2}\) with \(\Lambda = \omega_{\text{ce}}/\omega\), and are crossed in a fixed order; the remaining two, the helicon cutoff \(n_{\text{cut}} = k_{z}^{2}B_{0}/\omega\mu_{0}e\) and the coalescence density \(n_{c} = k_{z}^{2}B_{0}^{2}/4\mu_{0}m_{e}\omega^{2}\), depend on the antenna as well as the medium and stand in the fixed ratio \(n_{\text{cut}}/n_{c} = 4/\Lambda\), with \(n_{c}\) lying inside the operating range of a low-field source. The magnetisation condition \(\nu_{m} < \omega_{\text{ce}}\) is independent of density and is crossed in time by gas heating and depletion rather than by ionisation; until it is satisfied, no separated branches exist however dense the plasma. This is why the E--H--W sequence is a temporal sequence, and why neutral dynamics are part of the mechanism rather than a refinement of it.

Combining the bounded-mode resonance condition with the coalescence condition yields the ratio \(\mathcal{R} = n_{\text{res}}/n_{c} = \left( 4/\Lambda \right)\sqrt{1 + T^{2}/k_{z}^{2}}\), which separates bounded modes that resonate cleanly from those whose resonance is embedded in the coalesced regime, and which implies the sharp lower bound \(\omega_{\text{ce}} > 4\omega\) on the field required for any bounded mode to resonate before coalescence, i.e. a bound independent of density, power and working gas. For the reference case this places the dividing line between the third and fourth axial modes, so that the low-order modes an antenna of finite length excites most strongly are precisely those whose resonances are degraded.

The two-timescale analysis separates local from global adiabaticity, and the separation is large. The correction to the instantaneous constitutive relation is of relative order \(\left( \omega\tau_{n} \right)^{- 1}\) and never exceeds \(6 \times 10^{- 4}\), so the dielectric tensor may be evaluated with the instantaneous density throughout, with no memory kernel. The driven boundary-value problem, by contrast, has a response time \(2Q/\omega\) that becomes comparable to the density e-folding time near a resonance, and is strongly non-adiabatic for the first tens of microseconds. All of the non-adiabatic physics of ignition, e.g. the delayed and reduced resonance peak, the chirped ringing that follows it, the hysteresis of the mode jumps, resides in the global problem and not in the local dielectric response. The swept-resonance response is obtained in closed form and is governed by the single parameter \(\mathcal{S} = 4Q^{2}\left| d\omega_{\text{res}}/dt \right|/\omega^{2}\); mode conversion at the coalescence surface follows the Landau--Zener form with a coupling \(\left| \kappa_{\text{LZ}} \right| \propto \left( \nu/\omega \right)^{1/2}\), so that the conversion efficiency falls monotonically as the discharge becomes less collisional and its profile steepens. This predicts a maximum in fractional edge absorption partway through the transient, confirmed here at 94\% near \(n_{e} \sim 3 \times 10^{17}\) m$^{-3}$.

For the numerical treatment of the same problem, three consequences are practical and immediate: the permittivity formulation must be used rather than a conductivity, and the EMHD reduction cannot initialise a transient; the spatial resolution requirement is set by the TG skin depth, which varies by a factor of 25 during ignition; and the disagreement between a quasi-static parametric sweep and a genuine transient solve is not numerical error but a direct measurement of \(\epsilon_{\text{glob}}\), and should be used as such.

The principal limitation of the present theory is the cold-plasma approximation, which omits Landau damping. For the reference case \(\omega/k_{z}v_{\text{te}} \approx 3.7\), so kinetic damping is present but subdominant; it will become important for the TG branch in the late, less collisional transient, where the cold model will underestimate edge absorption. Extension to a warm or kinetic description of the TG branch, retaining the vacuum-regular structure that makes the present analysis possible, is the natural next step.

\appendix\section{Reduction to the quartic}
With \(\mathbf{E} \propto e^{i\mathbf{k} \cdot \mathbf{r}}\) and refractive index \(\mathbf{n} = \mathbf{k}/k_{0}\), the wave equation is \(\left\lbrack \mathbf{\text{nn}} - n^{2}\mathbb{I} + \mathbf{\varepsilon} \right\rbrack \cdot \mathbf{E} = 0\). Choosing \(\mathbf{n} = \left( n_{\bot},0,n_{\parallel} \right)\),
\begin{equation}
\begin{vmatrix}
S - n_{\parallel}^{2} & - iD & n_{\bot}n_{\parallel} \\
\text{iD} & S - n^{2} & 0 \\
n_{\bot}n_{\parallel} & 0 & P - n_{\bot}^{2} \\
\end{vmatrix} = 0
\end{equation}
Expanding along the second row gives
\begin{equation}
\left( S - n^{2} \right)\left\lbrack \left( S - n_{\parallel}^{2} \right)\left( P - n_{\bot}^{2} \right) - n_{\bot}^{2}n_{\parallel}^{2} \right\rbrack - D^{2}\left( P - n_{\bot}^{2} \right) = 0
\end{equation}
and collecting powers of \(n_{\bot}^{2}\) using \(n^{2} = n_{\bot}^{2} + n_{\parallel}^{2}\),
\begin{equation}
S\,n_{\bot}^{4} + \left\lbrack \left( S + P \right)\left( n_{\parallel}^{2} - S \right) + D^{2} \right\rbrack n_{\bot}^{2} + P\left\lbrack \left( n_{\parallel}^{2} - S \right)^{2} - D^{2} \right\rbrack = 0
\end{equation}
Multiplying by \(k_{0}^{4}\) and substituting \(k_{\bot} = k_{0}n_{\bot}\), \(k_{z} = k_{0}n_{\parallel}\) yields the dimensional form of Section 3.1. The vacuum check \(S = P = 1\), \(D = 0\) returns \(\left( n^{2} - 1 \right)^{2} = 0\).

\section{The electron-magnetohydrodynamic quadratic}
With \(\mathbf{J} = \nabla \times \mathbf{B}/\mu_{0}\) and \(\nabla \times \mathbf{E} = i\omega\mathbf{B}\), taking the curl of the generalised Ohm's law and imposing \(\nabla \times \mathbf{B} = \beta\mathbf{B}\), and using \(\nabla \times \left( \mathbf{B} \times \widehat{z} \right) = \partial_{z}\mathbf{B} = ik_{z}\mathbf{B}\) for a solenoidal field,
\begin{equation}
\frac{B_{0}\beta}{en_{e}\mu_{0}}\left( ik_{z}\mathbf{B} \right) + \frac{\eta_{\text{eff}}\beta^{2}}{\mu_{0}}\mathbf{B} = i\omega\mathbf{B}
\end{equation}
Dividing by \(i\mathbf{B}/\mu_{0}\) and using \(\eta_{\text{eff}}/i = - m_{e}\left( \omega + i\nu \right)/e^{2}n_{e}\),
\begin{equation}
\frac{k_{z}B_{0}}{en_{e}}\beta - \frac{m_{e}\left( \omega + i\nu \right)}{e^{2}n_{e}}\beta^{2} = \mu_{0}\omega
\end{equation}
which on multiplication by \(e^{2}n_{e}/m_{e}\widetilde{\omega}\) gives the quadratic of Section 3.6. Setting \(\nu = 0\), the small root recovers \(\beta k_{z} = \omega\mu_{0}n_{e}e/B_{0}\) and the large root \(\beta = k_{z}\omega_{\text{ce}}/\omega\).

\section{Derived quantities for the reference case}
\begin{table}[H]
\centering
\footnotesize
\setlength{\tabcolsep}{3pt}
\caption{Quantities derived from the reference case of Table 1, as used in the benchmarks of Section 8.4. The inputs themselves are not repeated here.}
\label{tab:7}
\begin{tabularx}{\textwidth}{>{\raggedright\arraybackslash}X>{\raggedright\arraybackslash}X}
\toprule
Quantity & Value \\
\midrule
\(\omega\) & 8.520 $\times$ 10$^{7}$ rad s$^{-1}$ \\
\(\kappa = \sqrt{k_{z}^{2} - k_{0}^{2}}\) & 31.42 m$^{-1}$ \\
\(T = 3.8317/a\) & 153.3 m$^{-1}$ \\
\(n_{1}\) & 2.281 $\times$ 10$^{12}$ m$^{-3}$ \\
\(n_{2}\) & 4.708 $\times$ 10$^{13}$ m$^{-3}$ \\
\(n_{3}\) & 9.720 $\times$ 10$^{14}$ m$^{-3}$ \\
\(n_{\text{cut}} = (4/\Lambda)n_{c}\) & 5.754 $\times$ 10$^{17}$ m$^{-3}$ (5.476 $\times$ 10$^{17}$ from full quartic) \\
\(n_{c}\) (\(k_{z} = 31.4\) m$^{-1}$) & 2.969 $\times$ 10$^{18}$ m$^{-3}$ \\
\(n_{\text{res}}\) (\(m = 1\), \(p = 4\)) & 2.865 $\times$ 10$^{18}$ m$^{-3}$ \\
\(k_{z}^{*}\) & 30.27 m$^{-1}$ (\(p^{*} = 3.85\)) \\
\(\sqrt{\omega_{\text{ce}}\omega_{\text{ci}}}/2\pi\) & 1.037 MHz \\
\(\tau_{n}\) (model) & 22 $\mu$s \\
\bottomrule
\end{tabularx}
\end{table}

\section{Nomenclature}
Symbols are listed in Latin then Greek alphabetical order. A descriptive subscript
is set upright and an index that takes values is set italic, so that \(n_{e}\) carries
the label "electron" whereas \(n_{s}\) runs over species.

\begin{table}[H]
\centering
\footnotesize
\setlength{\tabcolsep}{20pt}
\caption{Nomenclature: Latin symbols.}
\label{tab:8}
\begin{tabularx}{\textwidth}{l>{\raggedright\arraybackslash}X}
\toprule
Symbol & Meaning \\
\midrule
\(A\), \(A_{j}\) & slowly varying mode amplitude; WKB amplitude of branch \(j\) \\
\(\mathcal{A},\mathcal{B},\mathcal{C}\) & coefficients of the quartic in \(k_{\bot}^{2}\) \\
\(D\), \(D_{a}\) & gyrotropic element of the dielectric tensor; ambipolar diffusion coefficient \\
\(\mathbf{E}\), \(\widetilde{\mathbf{E}}\) & wave electric field and its slowly varying envelope \\
\(I_{m}\), \(J_{m}\) & modified Bessel and Bessel functions of the first kind, order \(m\) \\
\(\mathbf{J}\), \(\mathbf{J}_{a}\) & current density; antenna current density \\
\(k_{0}\), \(k_{\bot}\), \(k_{z}\) & vacuum, perpendicular and axial wavenumbers \\
\(k_{s}\), \(k_{w}\) & wavenumbers of the electron-magnetohydrodynamic quadratic \\
\(\mathcal{K}\) & \(k_{z}^{2}-k_{0}^{2}\) \\
\(K_{\text{iz}}\) & ionisation rate coefficient \\
\(L\), \(L_{n}\) & source length; density scale length \(|\nabla \ln n_{e}|^{-1}\) \\
\(\mathcal{L}\), \(\mathcal{M}\) & wave operator; response matrix of the constitutive relation \\
\(\mathcal{N}\) & wave action \(W/\omega\), the adiabatically conserved quantity \\
\(n_{c}\), \(n_{\text{cut}}\) & coalescence density; helicon cutoff density \((4/\Lambda)n_{c}\) \\
\(n_{\text{res}}\) & resonant density of a bounded eigenmode \\
\(p\), \(p^{*}\) & axial mode number, \(k_{z}=p\pi/L\); its critical value \\
\(p(r)\), \(P_{\text{abs}}\) & deposited power density; absorbed power \\
\(P\) & parallel element of the dielectric tensor \\
\(P_{\text{na}}\) & probability of failing to convert at the coalescence layer \\
\(Q\), \(Q_{\text{Helicon}}\) & quality factor; that of the helicon branch, \(\text{Re}\,\beta/2|\text{Im}\,\beta|\) \\
\(R_{p}\), \(\mathcal{R}\) & plasma loading resistance; resonance--coalescence ratio \(n_{\text{res}}/n_{c}\) \\
\(S\), \(\mathcal{S}\) & sum element of the dielectric tensor; sweep parameter \\
\(S_{\text{iz}}\), \(S_{\text{loss}}\) & ionisation source and loss terms \\
\(t\), \(\bar{t}\) & time; slow time \(\xi t\) of the two-timescale expansion \\
\(T\), \(T_{e}\) & perpendicular eigenvalue of the radial problem; electron temperature \\
\(\mathbf{v}_{e}\), \(v_{\text{te}}\) & electron fluid velocity; electron thermal speed \\
\bottomrule
\end{tabularx}
\end{table}

\clearpage
\begin{table}[H]
\centering
\footnotesize
\setlength{\tabcolsep}{20pt}
\caption{Nomenclature, continued: Greek symbols.}
\label{tab:9}
\begin{tabularx}{\textwidth}{l>{\raggedright\arraybackslash}X}
\toprule
Symbol & Meaning \\
\midrule
\(\alpha\), \(\gamma\) & sweep rate of the detuning; damping rate \(\omega/2Q\) \\
\(\beta\), \(\beta_{\pm}\) & total wavenumber \((k_{\bot}^{2}+k_{z}^{2})^{1/2}\); its two roots \\
\(\beta_{\text{Helicon}}\), \(\beta_{\text{TG}}\) & the helicon and Trivelpiece--Gould branches \\
\(\Gamma\) & Landau--Zener parameter governing branch conversion \\
\(\delta(t)\), \(\Delta\delta\) & instantaneous detuning; its shift at the response peak \\
\(\delta S,\delta P,\delta D\) & perturbations of the dielectric-tensor elements about vacuum \\
\(\delta k_{\bot}\) & splitting of the perpendicular wavenumber \\
\(\delta_{\nu}\), \(\delta_{\text{TG}}\) & collisionality \(\nu_{m}/\omega\); TG skin depth \\
\(\epsilon_{0}\), \(\mu_{0}\) & vacuum permittivity and permeability \\
\(\mathbf{\varepsilon}\), \(\mathbf{\varepsilon}_{H}\), \(\mathbf{\varepsilon}_{A}\) & dielectric tensor and its Hermitian and anti-Hermitian parts \\
\(\epsilon_{\text{loc}}\), \(\epsilon_{\text{glob}}\) & local and global adiabaticity parameters \\
\(\epsilon_{r}\) & radial WKB parameter \(1/k_{\bot}L_{n}\) \\
\(\zeta\) & size of a generic perturbation of a repeated root \\
\(\eta_{\text{eff}}\), \(\mathbf{\sigma}\) & effective resistivity; conductivity tensor \\
\(\kappa\), \(\kappa_{\text{LZ}}\) & vacuum evanescence rate \((k_{z}^{2}-k_{0}^{2})^{1/2}\); Landau--Zener coupling \\
\(\lambda_{z}\), \(\lambda_{\text{TG}}\) & axial and TG wavelengths \\
\(\Lambda\), \(\Lambda_{\nu}\) & magnetisation ratio \(\omega_{\text{ce}}/\omega\); magnetisation parameter \(\omega_{\text{ce}}/\nu_{m}\) \\
\(\nu\), \(\nu_{m}\), \(\nu_{\text{eff}}\) & collision frequency; momentum-transfer and effective values \\
\(\xi\) & two-timescale expansion parameter \((\omega\tau_{n})^{-1}\) \\
\(\tau_{n}\), \(\tau_{\text{cav}}\), \(\tau_{g}\) & ionisation, cavity-response \((2Q/\omega)\) and gas-depletion times \\
\(\Omega_{s}\) & signed gyrofrequency of species \(s\) \\
\(\omega\), \(\omega_{\text{res}}\) & drive angular frequency; resonant frequency \\
\(\omega_{\text{ce}}\), \(\omega_{\text{ci}}\) & electron and ion gyrofrequencies \\
\(\omega_{\text{pe}}\), \(\omega_{\text{pi}}\), \(\omega_{\text{p}s}\) & electron, ion and species-\(s\) plasma frequencies \\
\({\widetilde{\omega}}_{s}\) & collisionally shifted frequency \(\omega+i\nu_{s}\) \\
\bottomrule
\end{tabularx}
\end{table}

\begin{acknowledgments}
It is supported by National Natural Science Foundation of China (92271113, 12411540222, 12481540165), Natural Science Foundation Project of Chongqing (CSTB2025NSCQ-GPX0725), and ENN's Hydrogen--Boron Fusion Research Fund (2025ENNHB01-011). The author would like to also appreciate the enjoyable and pleasing environment in Architecture Library at Chongqing University, where this work was conducted. 
\end{acknowledgments}

\end{document}